\documentclass[aps,amsmath,amssymb,preprintnumbers,preprint,nofootinbib,a4paper,11pt]{article}
\pdfoutput=1
\usepackage{epsfig}
\usepackage{jheppub,undertilde}
\usepackage{amsthm}
\usepackage{graphicx}
\usepackage{graphbox}
\usepackage{xcolor}
\usepackage{graphicx,textcomp,float,slashed}
\usepackage{arydshln}
\usepackage{empheq}

\usepackage[normalem]{ulem}

\newcommand{\MeV}{\textrm{ MeV}}
\newcommand{\GeV}{\textrm{ GeV}}
 
\newcommand{\be}{\begin{equation}}
\newcommand{\ee}{\end{equation}}

\def\bsp#1\esp{\begin{split}#1\end{split}}
\def\bpm{\begin{pmatrix}}
\def\epm{\end{pmatrix}}

\newcommand{\bea}{\begin{eqnarray}}  
\newcommand{\eea}{\end{eqnarray}}  
 \def\bsp#1\esp{\begin{split}#1\end{split}}

\preprint{\begin{flushright} IFT-UAM/CSIC-19-115 \\
TTK-19-16 \\
KCL-PH-TH/2019-68
\end{flushright}}

\title{Probing Dark Matter freeze-in with long-lived particle signatures: 
MATHUSLA, HL-LHC and FCC-hh}

\author[1,3]{J.~M.~No}
\author[2,3]{, P.~Tunney}
\author[1]{and B.~Zaldivar}

\affiliation[1]{Departamento de Fisica Teorica and Instituto de Fisica Teorica, IFT-UAM/CSIC, Cantoblanco, 28049, Madrid, Spain.}
\affiliation[2]{Institute for Theoretical Particle Physics and Cosmology (TTK), RWTH Aachen University, D-52056 Aachen, Germany}
\affiliation[3]{Theoretical Particle Physics and Cosmology,
Department of Physics, King’s College London, London WC2R 2LS, United Kingdom}
\emailAdd{Josemiguel.no@uam.es}
\emailAdd{tunney@physik.rwth-aachen.de}
\emailAdd{bryan.zaldivarm@uam.es}

\abstract{
Collider searches for long-lived particles yield a promising avenue to probe the freeze-in production of Dark Matter via the decay of a parent particle.
We analyze the prospects of probing the parameter space of Dark Matter freeze-in from the decay of neutral parent particles at the LHC and beyond, taking as a case study a freeze-in Dark Matter scenario via the Standard Model Higgs.
We obtain the projected sensitivity of the proposed MATHUSLA surface detector (for MATHUSLA100 and MATHUSLA200 configurations) for long-lived particle searches to the freeze-in Dark Matter parameter space, and study its complementarity to searches by ATLAS and CMS at HL-LHC, as well as the interplay with constraints from Cosmology: Big-Bang Nucleosynthesis and Lyman-$\alpha$ forest observations. We then analyze the improvement in sensitivity that would come from a forward detector within a future 100 TeV $pp$-collider. 
In addition, we discuss several technical aspects of the present Dark Matter freeze-in scenario: the role of the electroweak phase transition; the inclusion of thermal masses, which have been previously disregarded in freeze-in from decay studies; the impact of $2\to 2$ scattering processes on the Dark Matter relic abundance; and the interplay between freeze-in and super-WIMP Dark Matter production mechanisms.}

\notoc

\begin{document}

\maketitle

\newpage

\tableofcontents
\section{Introduction}

While thermal freeze-out has long been considered the paradigm for Dark Matter (DM) production in the early Universe, experimental efforts over the last 
30 years have found no conclusive evidence of the existence of this type of DM.
At the same time, searches for DM in direct detection experiments~\cite{Cui:2017nnn,Aprile:2018dbl}, via indirect detection~\cite{Abdallah:2016ygi,Ahnen:2016qkx} and through collider probes (e.g.~at the LHC~\cite{Aaboud:2017phn,Sirunyan:2017jix}) yield at present strong bounds on the interaction strength between DM and the Standard Model (SM) particles. 

This provides motivation to consider alternative non-thermal DM production mechanisms in the early Universe. Among these, the {\sl freeze-in}
mechanism~\cite{McDonald:2001vt,Hall:2009bx} (see~\cite{Bernal:2017kxu} for a recent review) constitutes a very appealing possibility for its simplicity and the fact that it can be invoked in a wide range of well-motivated extensions of the SM (see e.g.~\cite{Covi:1999ty,Cheung:2011nn,Asaka:2005cn,Tsao:2017vtn,Goudelis:2018xqi,Garny:2018ali,Heeba:2018wtf,Heeba:2019jho,Darme:2019wpd,Hambye:2019dwd,Mambrini:2013iaa}). In freeze-in scenarios, DM particles are very feebly coupled to the thermal bath in the early Universe and never achieve thermal equilibrium, yet the coupling between DM and the thermal bath particles allows DM to be produced in decays and/or scatterings of bath particles\footnote{The inverse processes are absent due to the small DM abundance w.r.t.~equilibrium densities and to the feeble coupling between DM and the thermal bath.}. Through these processes, the DM abundance slowly increases towards equilibrium, however without ever reaching it. For renormalizable interactions between DM and the thermal bath, the production of DM is most efficient for temperatures $T$ around  ${\rm max}(m_A,m_{\rm DM})$, $m_A$ being the mass of the particle $A$ whose scatterings/decays produce DM\footnote{This is commonly known as {\sl infrared} freeze-in. For other possibilities, see e.g.~\cite{Mambrini:2013iaa,Elahi:2014fsa}.}. These processes shut-off (DM ``freezes-in") soon after $T$ drops below $m_A$ or $m_{\rm DM}$.

Given the feeble interactions between DM and the visible sector, freeze-in DM  candidates are naturally compatible with the current null results from DM experimental searches, and probing these scenarios may be more challenging than that of thermal DM candidates\footnote{See e.g.~\cite{Kahlhoefer:2018xxo}, although see \cite{Bernal:2015ova,Hambye:2018dpi,Heeba:2019jho} for recent phenomenological probes of freeze-in through a portal.}. Nevertheless, when freeze-in DM production in the early Universe proceeds via the decay of thermal bath particles~\cite{Hall:2009bx}, the corresponding feeble coupling would make these particles long-lived. Searches for long-lived particles (LLPs) at the LHC and beyond (see~\cite{Alimena:2019zri,Curtin:2018mvb}) have received renewed attention in recent years from the lack of conclusive signals in prompt LHC searches for physics beyond the SM, and provide a promising avenue for probing freeze-in DM scenarios~\cite{Co:2015pka,Hessler:2016kwm,Ghosh:2017vhe,Calibbi:2018fqf,Belanger:2018sti}.

In this work we analyze the prospects of probing the parameter space of DM freeze-in from the decay of long-lived neutral parent particles at the LHC and future colliders. As opposed to charged/coloured LLPs, which leave tracks in the LHC detectors and could be discovered via searches for heavy stable charged particles (see~\cite{Belanger:2018sti} for an analysis of DM freeze-in LHC signatures in these scenarios), neutral LLPs leave no trace in the 
ATLAS and CMS detectors until their decay. They constitute an ideal search objective for the proposed MATHUSLA surface detector~\cite{Curtin:2018mvb,Chou:2016lxi,Curtin:2017izq,Alpigiani:2018fgd} and other recent proposals for LLP detectors~\cite{Feng:2017uoz,Ariga:2018pin,Gligorov:2017nwh,Gligorov:2018vkc}. 

As a concrete model, we study a freeze-in DM scenario via the SM Higgs (a simple version of a supersymmetric Higgsino-Axino~\cite{Covi:2001nw,Co:2015pka,Co:2016fln} or Higgsino-singlino system, very similar to the so-called singlet-doublet DM scenario from~\cite{Calibbi:2018fqf,Calibbi:2015nha}). We
perform an analysis of the projected sensitivity of MATHUSLA in its 100 m $\times$ 100 m (MATHUSLA100) and 200 m $\times$ 200 m (MATHUSLA200) design proposals~\cite{Alpigiani:2018fgd}, to the freeze-in DM parameter space, and study its complementarity to LLP searches with the ATLAS and CMS detectors which use displaced vertices and missing transverse energy $E^{\rm miss}_T$~\cite{Calibbi:2018fqf}. We then analyze the improvement in sensitivity than would come from a forward detector within a future 100 TeV $pp$ - collider ({\sl FCC-hh}). 

In addition, we discuss in detail several technical aspects of the present DM freeze-in scenario (the latter three being in fact more general): the role of the breaking of electroweak (EW) symmetry in the early Universe (the EW phase transition) on the freeze-in production of DM; the inclusion of thermal masses on the computation of the DM relic density, previously disregarded in DM freeze-in studies (there have been however very recent efforts to include these effects as part of fully-consistent DM freeze-in analyses~\cite{Heeba:2019jho,Hambye:2019dwd,Darme:2019wpd}); the potential impact of $2\to 2$ scattering processes on setting the DM relic abundance and the interplay between freeze-in and super-WIMP DM production mechanisms (see also~\cite{Garny:2018ali} for a detailed analysis on the latter). Finally, we analyze the constraints from cosmology, namely from Big-Bang Nucleosynthesis (BBN) and from limits on the washout of small-scale structure via Lyman-$\alpha$ forest observations.

Our work is organized as follows: we give a review of the freeze-in DM production mechanism in section~\ref{sec:freezein_Review}. In section~\ref{subsec:freezeinHiggs} we discuss DM freeze-in via the SM Higgs, introduce the model and analyze the DM relic density including the impact of the EW phase transition, of including thermal masses in the calculation and of the super-WIMP DM production mechanism. In section~\ref{cosmology_Sec} we discuss the constraints from BBN and from Lyman-$\alpha$ forest observations. In section~\ref{MATHUSLA_Sec} we obtain the projected sensitivity of MATHUSLA to the DM freeze-in parameter space, and in section~\ref{Sec:ATLAS_CMS} we analyze its complementarity with ongoing LHC searches: mono-jet, disappearing tracks and displaced vertices + $E^{\rm miss}_T$. In section~\ref{100TeV_Sec} we derive the future prospects of an {\sl FCC-hh} 100 TeV collider with a forward detector, and finally we conclude in section~\ref{sec_conclusions}. Various technical details are confined to the appendices: Appendix~\ref{AppendixA} discusses $1\to 2$ vs $2\to 2$ processes regarding the DM relic density. Appendix~\ref{AppendixB} presents details of the recasting procedure for the ATLAS displaced vertices + $E^{\rm miss}_T$ search~\cite{Aaboud:2017iio}. Appendix~\ref{AppendixC} explicitly presents the NLO+NLL Higgsino pair production cross sections obtained with {\tt Resummino-2.0.1}~\cite{Fuks:2012qx,Fuks:2013vua} and used for our sensitivity estimates throughout the paper.   
   
\section{Dark Matter through the freeze-in mechanism: review}
\label{sec:freezein_Review}

We review here the main aspects of freeze-in production of DM through the decay 
of a parent particle $A$ in thermal equilibrium with the plasma~\cite{Hall:2009bx}
\be
A \; \rightarrow \; B_{\rm SM} \, \chi\ ,
\label{eq:FIprocess}
\ee
where $B_{\rm SM}$ is a SM state and $\chi$ is the DM candidate. Initially (after reheating at the end of inflation) the DM abundance is assumed 
to be negligibly small and subsequently increases continuously as the parent particle $A$ within the thermal bath decays into DM during the 
radiation-dominated era\footnote{In this work we assume a standard cosmological history; 
see~\cite{Co:2015pka,DEramo:2017ecx} for the impact of a modified thermal history of the Universe in freeze-in DM production.}. 
This DM production process is effective as long as the parent particle is relativistic. As the temperature of the radiation bath drops below the mass of $A$, 
the abundance of the parent particle becomes exponentially suppressed and the DM production process ceases to be effective. DM is then said to freeze-in, 
with most of the $\chi$ particles produced at temperatures $T \sim m_A/3$.
The DM number density $n_{\chi}$ evolves according to the Boltzmann equation~\cite{Hall:2009bx}
\be
\frac{d n_{\chi}}{dt} + 3 H\, n_{\chi} =  \Gamma_A  \, n_A^{\rm eq} \,\frac{K_1 (m_A/T)}{K_2(m_A/T)} \ ,
\label{eq:BoltzFI}
\ee
where $\Gamma_A$ is the decay rate producing DM in Eq.~\eqref{eq:FIprocess} and $K_{1,2}(x)$ are the first and second modified Bessel functions 
of the 2$^{\mathrm{nd}}$ kind. The parent particle equilibrium number density $n_A^{\rm eq}$ can be well-approximated by
\be
n_A^{\rm eq} \approx \frac{g_A\,\xi}{2 \pi^2} \, m_A^2 \,T \, K_2(m_A/T) \, ,
\ee
where $g_A$ counts the spin d.o.f. of $A$ and $\xi = 2$ if the decaying particle is not self-conjugate (otherwise $\xi = 1$). At high temperatures 
($T \gg m_A$), $n_A^{\rm eq} \sim T^3$ characteristic of a relativistic species, while it features the Maxwell-Boltzmann exponential suppression for $T \ll m_A$. 

The Boltzmann equation~\eqref{eq:BoltzFI} is general, with the details of the cosmological history entering through the Hubble parameter $H$ and 
the time vs temperature relation. 
Conservation of entropy $s=\frac{2\pi^2}{45}g_*^s T^3$ (with $g_*^s$ being the number of effective degrees of freedom contributing to the entropy density) in general makes it convenient to track the evolution of the DM population by defining the comoving number density (yield), $Y_{\chi}\equiv n_{\chi}/s$, which is dimensionless, and on the other hand allows us to relate time and temperature 
as $dt=-dT/(\bar H T)$, where (see e.g.~\cite{Belanger:2018mqt})
\be 
\bar H\equiv \dfrac{H}{1+\frac{1}{3}\frac{d\ln g^s_*}{d\ln T}}\,.
\ee 
For a radiation-dominated universe the Hubble parameter is $H\simeq 1.66\sqrt{g_*}\,T^2/M_{\rm Pl}$, where $g_*$ is the number of effective 
relativistic degrees of freedom contributing to the energy density, and $M_{\mathrm{Pl}} = 1.2 \times 10^{19}$ GeV is the Planck mass. Describing the evolution in terms of $x \equiv m_A/T$ and 
the yield $Y_{\chi}$, the Boltzmann equation~\eqref{eq:BoltzFI} reads
\be
\frac{d Y_{\chi}}{d x} = \frac{\Gamma_A}{H \, x} \, Y_{A}^{\rm eq}(x) \, \frac{K_1(x)}{K_2(x)} \ ,
\ee
with $Y_A^{\rm eq} = n_A^{\rm eq} / s$. Assuming that at very high temperatures $T_R \gg m_A$ the abundance 
of $\chi$ is vanishing, this equation can be integrated to give
\be 
\label{Ychi_1}
Y_\chi\approx \dfrac{45\,g_A \,\xi}{(1.66)\,8\,\pi^4}\dfrac{M_{\rm Pl}\,
\Gamma_A}{m_A^2}\int_{x_{\rm min}}^{x_{\rm max}}\dfrac{1+\frac{1}{3}\frac{d\ln g^s_*(x)}{d\ln x}}{g_*^s(x)\sqrt{g_*(x)}} K_1(x)\,x^3 dx\, ,
\ee
with $x_{\rm min} = m_A/T_R \ll 1$ and $x_{\rm max} \to \infty$.
Far from phase transitions where the number of d.o.f.~in the plasma could potentially change abruptly, the DM yield is to a good approximation
\bea 
\label{Ychi_2}
Y_\chi&\approx& \dfrac{135\, g_A\,\xi\, M_{\rm Pl}\,\Gamma_A}{16\,\pi^3\,m_A^2 }\frac{1}{1.66\,g_*^s(m_A/3)\sqrt{g_*(m_A/3)}}\,.
\eea
Requiring then the relic abundance of $\chi$ 
\be
\label{Omegachi_1}
\Omega_{\chi} h = \frac{m_{\chi} \, Y_{\chi}}{\rho_c/s_0}\,,
\ee
(with $\rho_c/s_0 = 3.6 \times 10^{-9}$ GeV the critical energy density over the entropy density today) to match the observed DM relic 
abundance $\Omega_{\mathrm{DM}} h = 0.12$, we can obtain an estimate of the corresponding decay length $c\tau_A$ of the parent particle:
\be 
\label{lifetime_A}
c\tau_A\sim 3.6 \text{ m } 
\left( \frac{g_A\,\xi}{4} \right)  \left(\frac{106.75}{g_*}\right)^{3/2} 
\left(\frac{m_\chi}{100~{\rm keV}}\right) \left(\frac{300 \, \GeV}{m_A}\right)^2~,
\ee 
where we have assumed for simplicity $g_*^s(x)=g_*(x)$ at $T=m_A/3$ (approximately the freeze-in temperature, see \cite{Belanger:2018mqt}). 
We note the macroscopic lifetime for the parent particle $A$, which according to Eq.~\eqref{lifetime_A} increases linearly with the DM mass $m_{\chi}$
as a result of requiring  $\Omega_{\chi} h = 0.12$.

\section{Freeze-in (from decay) via the Standard Model Higgs}
\label{subsec:freezeinHiggs}

In the present work we concentrate on freeze-in DM scenarios where the parent particle $A$ in Eq.~\eqref{eq:FIprocess} is 
neutral, which are significantly more challenging to probe experimentally than those with a charged parent 
particle\footnote{For a detailed phenomenological study of freeze-in scenarios where the parent particle is electrically charged and/or coloured, see~\cite{Belanger:2018sti}.}, 
particularly if the decay length of $A$ is large, $c\tau_A \gg 1$ m. In the following we consider 
$B_{\mathrm{SM}}$ in Eq.~\eqref{eq:FIprocess} to be the SM Higgs~\cite{Calibbi:2018fqf,Co:2015pka,Co:2016fln}. 

\subsection{The model}

A simple model which features the SM Higgs as $B_{\mathrm{SM}}$ consists on adding to the SM a Dirac fermion $\chi$, singlet under the SM gauge group, 
and an $SU(2)_L$ Dirac fermion doublet $\Psi$ with hypercharge $1/2$
\begin{equation}
\label{psi_field}
\Psi = \left( 
\begin{array}{c}
\psi^{+}\\
\psi^0
\end{array}
\right)
\end{equation}
such that the Lagrangian reads:
\begin{equation}
\label{eq:DMSM_FI}
\mathcal{L} = \mathcal{L}_{\mathrm{SM}} +  
i \,\bar{\chi} \gamma^{\mu} \partial_{\mu} \chi + i \,\bar{\Psi} \gamma^{\mu} D_{\mu} \Psi - m_s \,\bar{\chi}\chi - m_D \bar{\Psi}\Psi
- y_{\chi} \, \bar{\Psi} H \chi + h.c. 
\end{equation}
This model is very similar to the singlet-doublet DM model considered in~\cite{Calibbi:2015nha,Calibbi:2018fqf}, and could be regarded as a simplified 
version of the Higgsino-Axino system studied in~\cite{Covi:1999ty,Covi:2001nw,Co:2015pka,Co:2016fln}, 
or an analog of a would-be feebly interacting Higgsino-Bino or Higgsino-Singlino system, bearing in mind that here $\chi$ and $\Psi$ are Dirac fermions. 

\vspace{2mm}

\begin{figure}[h]
\begin{center}
\includegraphics[width=0.49\textwidth]{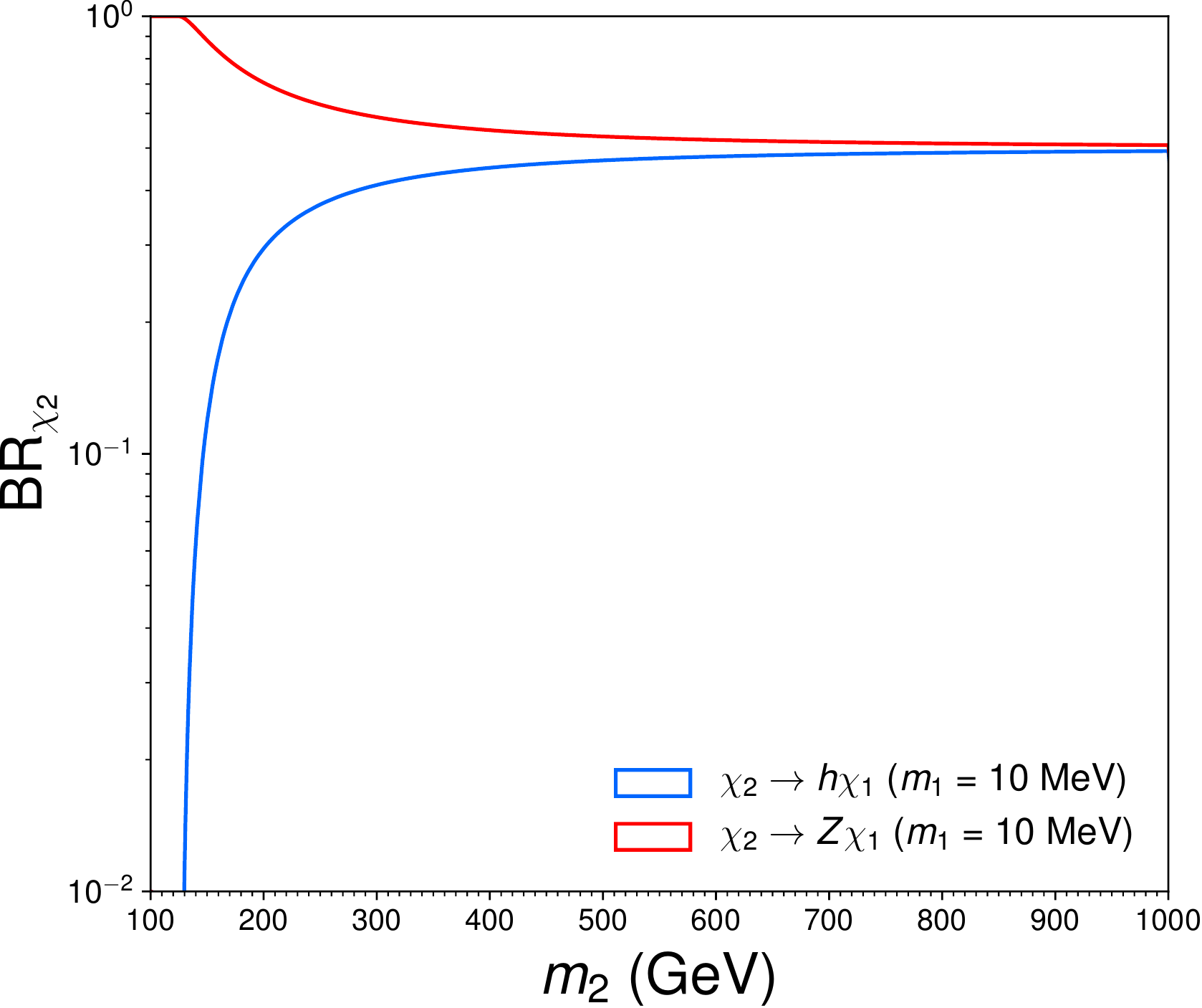}
\includegraphics[width=0.49\textwidth]{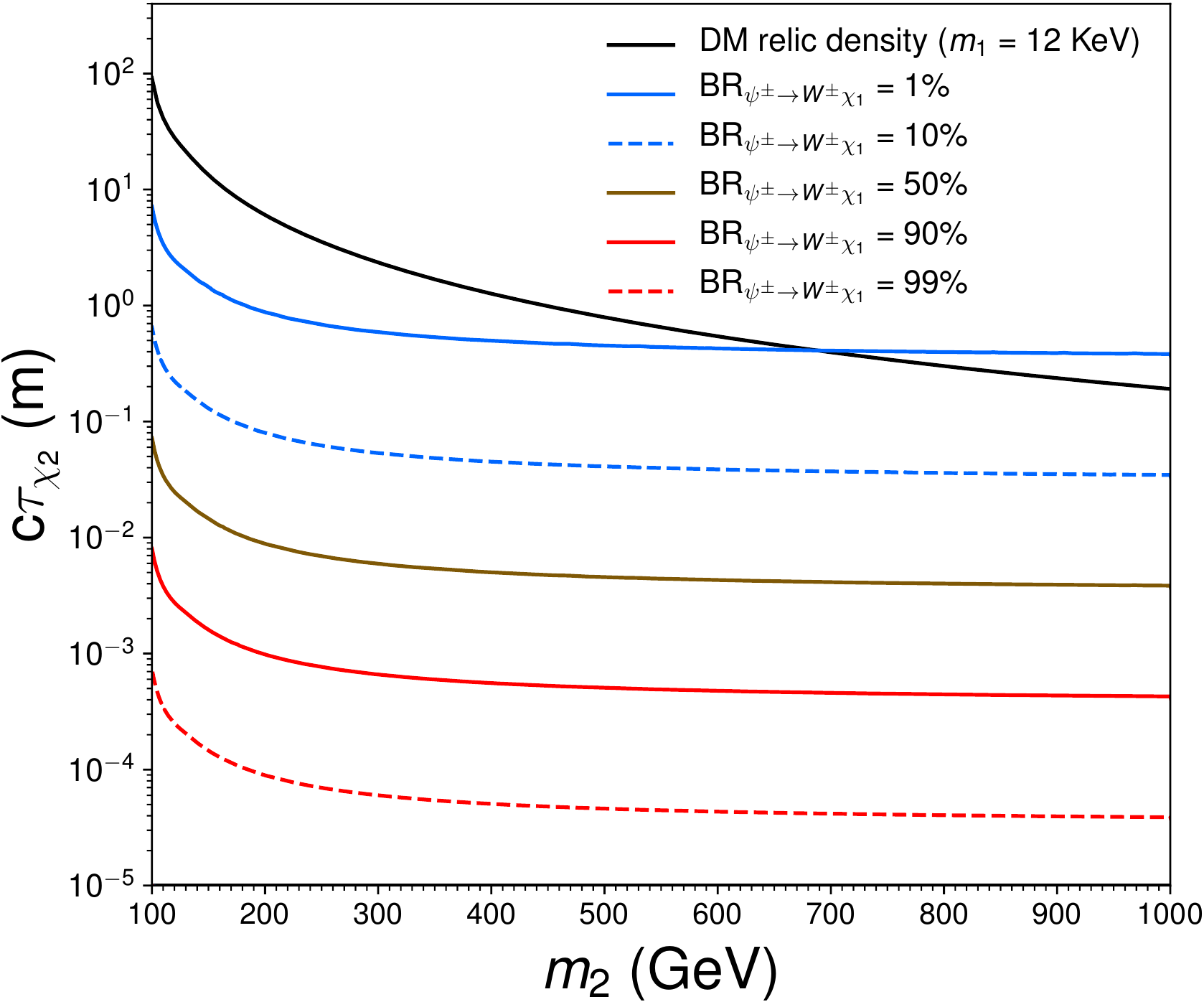}

\vspace{-2mm}
\caption{\small LEFT: Branching fractions for $\chi_2 \to h \chi_1$ (blue) and $\chi_2 \to Z \chi_1$ (red) as a function 
of $m_2$ for $m_1 = 10$ MeV. RIGHT: Branching ratio BR($\psi^{\pm} \to W^{\pm} \chi_1$) values in the ($m_2$, $c\tau_{\chi_2}$ plane for a DM mass $m_{1} = 12$ KeV. 
The solid black line corresponds to  $\Omega_{\chi} h = 0.12$ for $m_{1} = 12$ KeV as computed with {\tt micrOMEGAs5.0}~\cite{Belanger:2018mqt} (assuming an 
EW phase transition temperature $T_{\mathrm{EW}} = 160$ GeV, see the discussion in Section~\ref{subsec:DM_Relic_EWSB}).}
\label{BR_chi2}
\end{center}

\vspace{-3mm}

\end{figure}

The neutral particles $\chi$ and $\psi^0$ mix after EW symmetry breaking due to the presence of the $y_{\chi}$ coupling, giving rise to mass eigenstates $\chi_{1}$ (mostly singlet-like) and $\chi_{2}$ (mostly doublet-like), their respective masses being $m_1$ and $m_2$. We consider $m_2 > m_1$ and the coupling $y_\chi \ll 1$, as needed for the freeze-in mechanism to yield the correct DM relic density. In this limit we have $m_1 \simeq m_s$ and $m_2 \simeq m_D$ in Eq.~\eqref{eq:DMSM_FI} ($m_2 > m_1$ is then achieved by setting $m_{D}  > m_{s}$). The singlet-doublet mixing is simply given by
\begin{equation}
\mathrm{sin}\, \theta \simeq \frac{y_{\chi} v}{\sqrt{2} (m_2 - m_1)}
\end{equation}
valid in the limit $m_2 - m_1 \gg y_{\chi} v$, 
with $v = 246$ GeV the SM Higgs vev.
After EW symmetry breaking, the interactions of the DM candidate $\chi_{1}$ are then given by $h-\chi_2-\chi_1$, directly from the Yukawa term in Eq.~\eqref{eq:DMSM_FI}, and $Z-\chi_2-\chi_1$, $W^{\pm}-\psi^{\mp}-\chi_1$ from the singlet-doublet mixing induced by the Yukawa term after EW symmetry breaking. When kinematically possible, the decay widths for $\chi_2 \to h \chi_1$, $\chi_2 \to Z \chi_1$ and $\psi^{\pm} \to W^{\pm} \chi_1$ are given by
\begin{eqnarray}
 \Gamma (\chi_2 \to h \chi_1) &=& \frac{y_{\chi}^2}{32\,\pi\,\, m_{2}^3} \left[(m_2+m_1)^2 - m_h^2\right] \lambda(m_2, m_1, m_h) \nonumber \\
 \label{decayFIDM_1}
 \Gamma (\chi_2 \to Z \chi_1) &=&  \frac{y_{\chi}^2}{32\,\pi} 
 \frac{\left[(m_2 - m_1)^2 - m_Z^2\right]\left[(m_2+m_1)^2 + 2 m_Z^2\right]}{m_{2}^3\,\,(m_2-m_1)^2} \lambda(m_2, m_1, m_Z) \\
 \Gamma (\psi^{\pm} \to W^{\pm} \chi_1) &=&  \frac{y_{\chi}^2}{32\,\pi} 
 \frac{\left[(m_{\psi} - m_1)^2 - m_W^2\right]\left[(m_{\psi}+m_1)^2 + 2 m_W^2\right]}{m_{\psi}^3\,\,(m_{\psi}-m_1)^2} \lambda(m_{\psi}, m_1, m_W) \nonumber
\end{eqnarray}
%
with $\lambda(x, y, z) = \sqrt{x^4 + y^4 + z^4 - 2 x^2 y^2 - 2 x^2 z^2 - 2 y^2 z^2}$. In Figure~\ref{BR_chi2} (left) we show the branching fractions of $\chi_2$ as a function of $m_2$ for a benchmark\footnote{Compared to 
the results in~\cite{Calibbi:2018fqf}, ours correspond to the choice tan\,$\theta = 1$ in~\cite{Calibbi:2018fqf}.} $m_1 = 10$ MeV, 
highlighting that for $m_2 \lesssim 400$ GeV the decay $\chi_2 \to Z \chi_1$ starts to dominate over $\chi_2 \to h \chi_1$. For $m_2 - m_1 \to m_h$ 
the branching fraction into $h \chi_1$ becomes negligible, while for $m_2 - m_1 \gg m_h$ both branching fractions approach $50\%$. 

In addition, loops of EW gauge bosons induce a radiative mass splitting between the charged state $\psi^{\pm}$ and the (mostly doublet-like) 
neutral state $\chi_2$~\cite{Thomas:1998wy,Cirelli:2009uv}. The value of this mass splitting $\delta m$ is given by
\begin{equation}
\label{deltam_charged}
\delta m = m_{\psi^{\pm}} - m_2 =  \frac{g^2\, m_Z}{ 32 \pi^2}\, \mathrm{sin}^2\theta_W \,f \left(\frac{m_z}{m_2}\right)\, ,
\end{equation}
with $g$ the $SU(2)_L$ gauge coupling, $\theta_W$ the Weinberg angle and $f(m_Z/m_2)$ given by
\begin{equation}
f(r) =  2 r^3 {\rm log} \,r - 2 r + (r^2 +2) \sqrt{r^2 - 4}\, {\rm log}\left( \frac{r^2 - 2 - r \sqrt{r^2 -4}}{2} \right)\, .
\end{equation}
For $m_2 \in [100, 1000]$ GeV the range of mass splitting values is $\delta m \in [260, 340]$ MeV. For such splittings the decay mode 
$\psi^{\pm} \to \pi^{\pm} \chi_2$ completely dominates over the decay involving charged leptons $\psi^{\pm} \to \ell^{\pm} \nu \chi_2$~\cite{Thomas:1998wy}, with the width
$\Gamma(\psi^{\pm} \to \pi^{\pm} \chi_2)$ given by 
\begin{equation}
\label{Gamma_charged}
\Gamma (\psi^{\pm} \to \pi^{\pm} \chi_2) =  \frac{\mathrm{cos}^2\theta_c}{ 2\, \pi \,v^4}\, f^2_{\pi} \, \delta m^3 \sqrt{1 -  \frac{m^2_{\pi}}{\delta m^2}}\, ,
\end{equation}
with $f_{\pi} \simeq 130$ MeV and $\theta_c$ the Cabibbo angle. For the very small values of the singlet-doublet mixing relevant for DM freeze-in, 
the decay width from Eq.~\eqref{Gamma_charged} is generally much
larger than $\Gamma(\psi^{\pm} \to W^{\pm} \chi_1)$ and yields the dominant decay mode for $\psi^{\pm}$. This can be seen explicitly in Figure~\ref{BR_chi2} (right), 
where we show the branching ratio BR($\psi^{\pm} \to W^{\pm} \chi_1$) as a function of $m_2 \simeq m_{\psi^{\pm}}$ and the $\chi_2$ decay length $c\tau_{\chi_2}$ (which can be directly related to $\Gamma(\psi^{\pm} \to W^{\pm} \chi_1)$ by means of Eq.~\eqref{decayFIDM_1}) for a benchmark DM mass $m_{1} = 12$ KeV. We also show in Figure~\ref{BR_chi2} (right) the DM relic density curve for $m_{1} = 12$ KeV 
as computed with {\tt micrOMEGAs5.0}~\cite{Belanger:2018mqt} (see Section~\ref{subsec:DM_Relic_EWSB} for details); since roughly all parameter space below this curve is excluded by the Lyman-$\alpha$ bound from Cosmology (see Section~\ref{cosmology_Sec}), Figure~\ref{BR_chi2} (right) highlights that, in order to obtain the observed DM relic abundance, BR($\psi^{\pm} \to W^{\pm} \chi_1$) $\ll 0.1$ for $m_2 < 1$ TeV is needed in the present freeze-in scenario.

\subsection{DM relic density: the role of EW symmetry breaking \& thermal masses}
\label{subsec:DM_Relic_EWSB}

As discussed in Section~\ref{sec:freezein_Review}, the DM relic abundance is obtained via slow $\chi_1$ production from the decays of the parent particles $\chi_2$, $\psi^{\pm}$ present in the thermal bath during the radiation-dominated era. For the computation of the DM relic density, we have used 
{\tt micrOMEGAs5.0}~\cite{Belanger:2018mqt} with the additional implementation of several important features, which we detail in the following.

Prior to EW symmetry breaking in the early Universe (the EW phase transition), the interactions of DM with the EW gauge bosons are absent\footnote{We are indebted to Thomas Konstandin for discussions on this point.}, and the states $\psi^0$, $\psi^{+}$ in~\eqref{psi_field} decay via the Yukawa interaction $y_{\chi}$ into the SM Higgs doublet field components and the DM candidate, $\Psi \to H \chi$. 
The correct evaluation of DM production in the EW symmetric phase requires the inclusion of the temperature-dependent (thermal) mass $\Pi(T)$ of the Higgs doublet, given by~\cite{Carrington:1991hz,Comelli:1996vm} (see also~\cite{Katz:2014bha})
\be
\label{thermal_mass_Higgs}
\Pi^2_H(T) = \left[\frac{3\,g^2}{16} +  \frac{g'^2}{16} + \frac{y_t^2}{4} + 
\frac{\lambda}{2}\right]\,T^2 \simeq (0.631\,T)^2\, ,
\ee
with $g$, $g'$ the gauge couplings for $SU(2)_{L}$ and $U(1)_{Y}$, $y_t$ the top quark Yukawa coupling and $\lambda$ the Higgs quartic coupling. On the other hand, since the parent particle is in equilibrium with the SM, it will also acquire a thermal mass (see e.g.~\cite{Rychkov:2007uq,Hambye:2017elz})\footnote{Here we use the thermal mass of a Higgsino (or analogously, that of a lepton $SU(2)_{L}$ doublet with hypercharge $Y = 1/2$). We note that the DM coupling of $\Psi$ may be safely neglected for this purpose.
} $\Pi_\Psi(T)$:
\be
\label{thermal_mass_Chi}
\Pi^2_\Psi(T) = \left[\frac{3\,g^2}{16} +  \frac{g'^2}{16}\right]\,T^2 \, ,
\ee
such that before the EW phase transition we consider the mass of $\Psi$ to be
\be 
m_{\Psi}(T) = m_D + \Pi_\Psi(T) \approx m_D + 0.293\,T~.
\label{thermal_mass_PSi}
\ee
The temperature dependence from both $\Pi_H(T)$ and $\Pi_\Psi(T)$ will then be inherited by the parent decay width in the EW symmetric phase\footnote{With the corresponding replacements $m_h = \Pi_H(T)$ and $m_2 \to m_2 + \Pi_\Psi(T)$. Note that $\Gamma_{\Psi\to H\chi}(T)$ approximately corresponds to four times the value of $\Gamma (\chi_2 \to h \chi_1)$ due to the number of degrees of freedom inside the $\Psi$ and SM Higgs doublets.}
\be
\Gamma_A(T>T_{\rm EW}) = \Gamma_{\Psi\to H\chi}(T)~,
\label{Gamma_bEWSB}
\ee
with $T_{\mathrm{EW}}$ the EW phase transition temperature, corresponding to $T_{\mathrm{EW}} \approx 160~\mathrm{GeV}$ in the SM~\cite{DOnofrio:2014rug,DOnofrio:2015gop}. Note that the thermal mass of the parent particle $\Psi$ is smaller than the thermal mass of the Higgs, so that for $ T \gg m_D \, (\simeq m_2)$ the decay $\Psi \to H \chi$ will not be kinematically open. However, for lower temperatures (yet above $T_{\mathrm{EW}}$) the decay may be open due to the presence of the bare mass $m_D$ in Eq.~\eqref{thermal_mass_PSi}.

\vspace{2mm}

After the EW phase transition the decays $\chi_2 \to Z \chi_1$ and $\psi^{\pm} \to W^{\pm} \chi_1$ become possible and source DM production together with the decay $\chi_2 \to h \chi_1$. The decay width $\Gamma_A$ of the parent particle(s) responsible for DM freeze-in is in this case given by 
\be
\label{Gamma_aEWSB}
\Gamma_A (T < T_{\rm EW}) = \Gamma (\chi_2 \to h \chi_1) +  \Gamma (\chi_2 \to Z \chi_1) + \Gamma (\psi^{\pm} \to W^{\pm} \chi_1)\, ,
\ee
where for simplicity we do not consider the temperature dependence of the Higgs vev after EW symmetry breaking and also neglect thermal corrections for temperatures below the EW phase transition.

\begin{figure}[h]
\begin{center}
\includegraphics[width=0.85\textwidth]{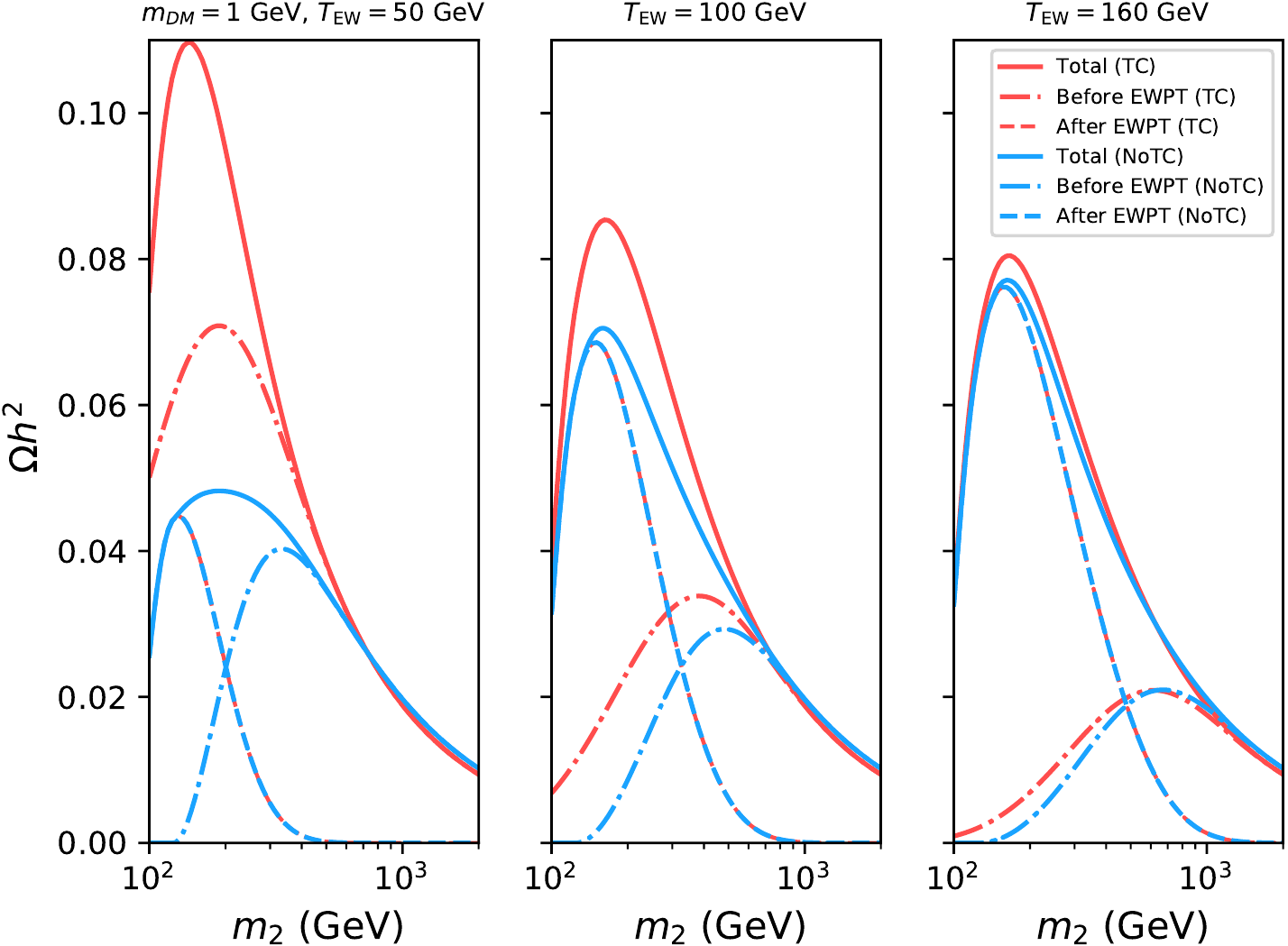}

\vspace{-2mm}

\caption{\small DM Relic density $\Omega h^2$ as a function of $m_2$, for $m_1 = 1$ GeV, $y_{\chi} = 10^{-10}$ and $T_{\mathrm{EW}} = 50$ GeV (left panel), $T_{\mathrm{EW}} = 100$ GeV (middle panel), $T_{\mathrm{EW}} = 160$ GeV (right panel). Red lines show the value of $\Omega h^2$ from DM production including thermal corrections: before the EW Phase Transition (EWPT, dot-dashed line), after the EWPT (dashed line), and the total amount (solid line). Light-blue lines show the corresponding value of $\Omega h^2$ disregarding thermal corrections: from DM production before the EWPT (dot-dashed), after the EWPT (dashed), and total (solid).}
\label{fig:thermal_masses_0}
\end{center}
\end{figure}

The temperature dependence of the freeze-in decay processes 
from Eqs.~\eqref{Gamma_bEWSB} and \eqref{Gamma_aEWSB} has been implemented in a modified version of {\tt micrOMEGAs5.0}, and in the remainder of this work we consider the range between $T_{\mathrm{EW}} = 160~\mathrm{GeV}$ and $T_{\mathrm{EW}} = 50~\mathrm{GeV}$ (the latter corresponding to a generic benchmark for a strongly super-cooled first order EW phase transition) when discussing/computing the freeze-in DM relic density.

\vspace{2mm}

Both the EW phase transition and the inclusion of thermal masses in the calculation can then have a notable influence on the DM freeze-in process\footnote{The 
potential influence of the EW phase transition on freeze-in processes has already been noted and discussed, albeit for different freeze-in scenarios to the freeze-in from decay $A \to B_{\mathrm{SM}} \chi$ considered in this work, in Refs.~\cite{Heeba:2018wtf,Heeba:2019jho,Darme:2019wpd,Lebedev:2019ton}.}.
In particular, the DM production prior to the EW phase transition can be strongly affected by thermal corrections. The production of DM after the EW phase transition is however not expected to be significantly affected, and as discussed above we are for simplicity neglecting here the thermal effects below $T_{\mathrm{EW}}$. It becomes clear that thermal effects will be more important when the contribution to the total DM relic density from DM production for $T > T_{\mathrm{EW}}$ is sizable.  
In Figure~\ref{fig:thermal_masses_0} we 
show the DM relic density with the inclusion of thermal corrections $\Omega h^2_{\mathrm{TC}}$ and the corresponding DM relic density when these are 
disregarded\footnote{Here thermal corrections are disregarded by setting $\Gamma_{\Psi\to H\chi} \simeq 4 \times \left.\Gamma (\chi_2 \to h \chi_1)\right|_{m_h = 125 \,\mathrm{GeV}}$.}
$\Omega h^2_{\mathrm{NoTC}}$, as well as the partial contributions before and after the EW phase transition, as a function of $m_2$ for $m_1 = 1$ GeV and $T_{\mathrm{EW}} = 50~\mathrm{GeV}$ (left panel), $T_{\mathrm{EW}} = 100~\mathrm{GeV}$ (middle panel) and $T_{\mathrm{EW}} = 160~\mathrm{GeV}$ (right panel). 
\begin{figure}[h]
\begin{center}

\includegraphics[width=0.80\textwidth]{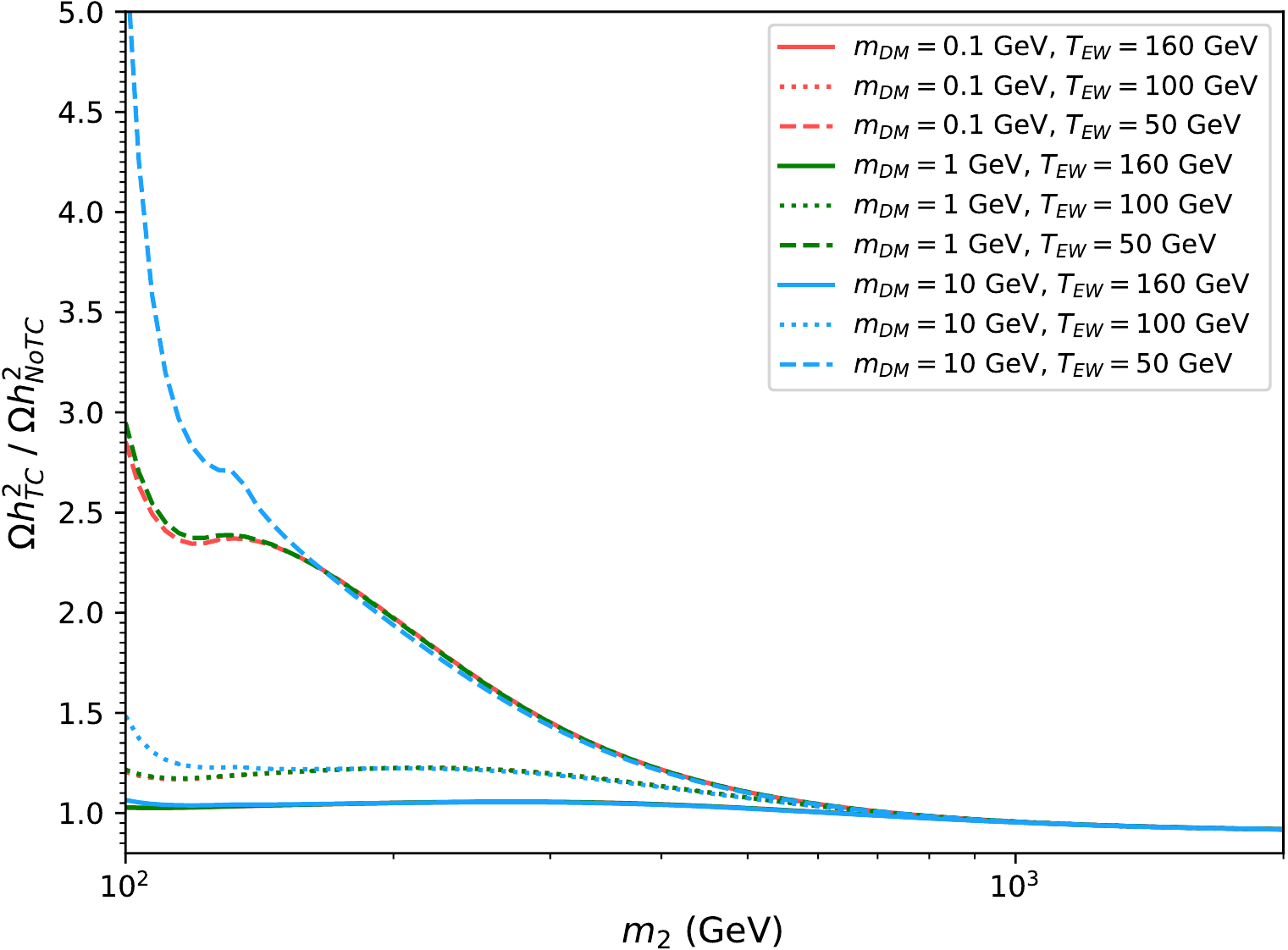}

\vspace{-2mm}
\caption{\small Ratio $\Omega h^2_{\mathrm{TC}}/\Omega h^2_{\mathrm{NoTC}}$ as a function of $m_2$ for a DM mass $m_1 = 0.1$ GeV (red), $m_1 = 1$ GeV (green) and $m_1 = 10$ GeV (light-blue), respectively for $T_{\mathrm{EW}} = 160$ GeV (solid line), $T_{\mathrm{EW}} = 100$ GeV (dotted line) and $T_{\mathrm{EW}} = 50$ GeV (dashed line).}
\label{fig:thermal_masses}
\end{center}
\end{figure}

As seen in Figure~\ref{fig:thermal_masses_0}, for $m_2 \gg m_h + m_1,\,T_{\mathrm{EW}}$ the inclusion of thermal masses yields a very mild suppression $\Omega h^2_{\mathrm{TC}}/\Omega h^2_{\mathrm{NoTC}} \sim 0.9$. 
In this case, DM freezes-in while the EW symmetry is restored, and the 
broken phase decays $\chi_2 \to Z \chi_1$ and $\psi^{\pm} \to W^{\pm} \chi_1$ do not contribute to DM production in the early Universe. In this regime we find the DM relic abundance to be rather insensitive to the value of $T_{\mathrm{EW}}$.
On the other hand, Figure~\ref{fig:thermal_masses_0} shows that 
the DM production in the EW broken phase starts becoming appreciable 
for $m_2 \lesssim 6 \,T_{\mathrm{EW}}$, and for yet smaller $m_2$ the DM production in the broken phase quickly becomes dominant when $T_{\mathrm{EW}} \gtrsim m_h$. 
In contrast, for $T_{\mathrm{EW}} \lesssim m_h$ the DM production prior to the EW phase transition is still very important for smaller $m_2$ masses, and it is in this region where the inclusion of the thermal corrections yields the largest effect, providing a significant enhancement to the DM relic density compared to the value neglecting thermal corrections. 
The difference in behaviour of the two regimes ($T_{\rm EW}$ much above or below the Higgs mass) is due to the fact that the latter case allows for the decay process prior to the EW phase transition to be kinematically open for a longer period. All this may be seen explicitly in Figure~\ref{fig:thermal_masses} where $\Omega h^2_{\mathrm{TC}}/\Omega h^2_{\mathrm{NoTC}}$ is shown as a function of $m_2$ for several values of $m_1$ and $T_{\mathrm{EW}}$. Note that below $m_2\lesssim m_h$ the enhancement to the DM relic abundance coming from thermal corrections is significantly larger for $m_{\rm DM}=10$ GeV than for $m_{\rm DM}=1$ GeV or less. This is due to phase space arguments, the parent decay being open or closed in the EW symmetric phase depending on the inclusion of thermal corrections or not, mostly for the lowest $T_{\rm EW}$ considered.

\vspace{2mm}

Finally, we stress that in the above discussion, and throughout the general discussion of freeze-in from decay in~\cite{Hall:2009bx}, outlined in Section~\ref{sec:freezein_Review}, 
it has been implicit that ($1 \to 2$) decays are the dominant processes producing DM. We nevertheless note that there are contributions to DM production at the same order in the feeble coupling $y_{\chi}$ from $2 \to 2$ scattering processes such as $q \,\chi_2 \to q\, \chi_1$ (with $q$ a SM quark) mediated by a Higgs boson in the $t$-channel. In a similar manner, for $T < T_{\mathrm{EW}}$ analogous $q \, \chi_2 \to q\, \chi_1$ and $q\, \chi^{\pm} \to q\,' \, \chi_1$ scattering processes with a $Z$ and $W$ boson in the $t$-channel become possible. 
These processes are discussed in detail in Appendix~\ref{AppendixA}, where we find that decays $A \to B_{\mathrm{SM}}\, \chi$
completely dominate over scatterings $X_{\rm SM} \, A \to Y_{\rm SM} \, \chi$ (with $X_{\rm SM}$ and $Y_{\rm SM}$ SM particles, and $B_{\rm SM}$ in the $t$-channel of the scattering process) except when the former are strongly phase-space suppressed, {\it i.e.}~$m_{B_{\mathrm{SM}}} + m_{\chi} \to m_A$. Scatterings also become dominant in the region $ m_{B_{\mathrm{SM}}} + m_{\chi} >  m_A > m_\chi$ where $1 \to 2$ decays are kinematically forbidden and instead $A$ decays\footnote{We assume that $B_{\rm SM}$ is itself unstable, otherwise for $ m_{B_{\mathrm{SM}}} + m_{\chi} >  m_A$ the state $A$ becomes stable.} via a 3-body process.
In our model, the latter situation occurs for $m_2 - m_1 < m_W$, when all the decays of $\chi_2$ and $\psi^{\pm}$ that produce DM are 3-body (e.g. $\chi_2 \to b \bar{b}\, \chi_1$ through an off-shell Higgs boson).
Since we do not consider these parameter space regions in the present work due to experimental constraints (e.g.~the region $m_{\psi^{\pm}} < m_W$ is strongly constrained by LEP~\cite{LEPSUSYWG,Egana-Ugrinovic:2018roi}), $2 \to 2$ scattering processes can be safely disregarded in our scenario.

\subsection{Super-WIMP contribution to the DM relic abundance}
\label{sec:DM_Relic_SW}

\begin{figure}[t]
\begin{center}
\includegraphics[width=0.7\textwidth]{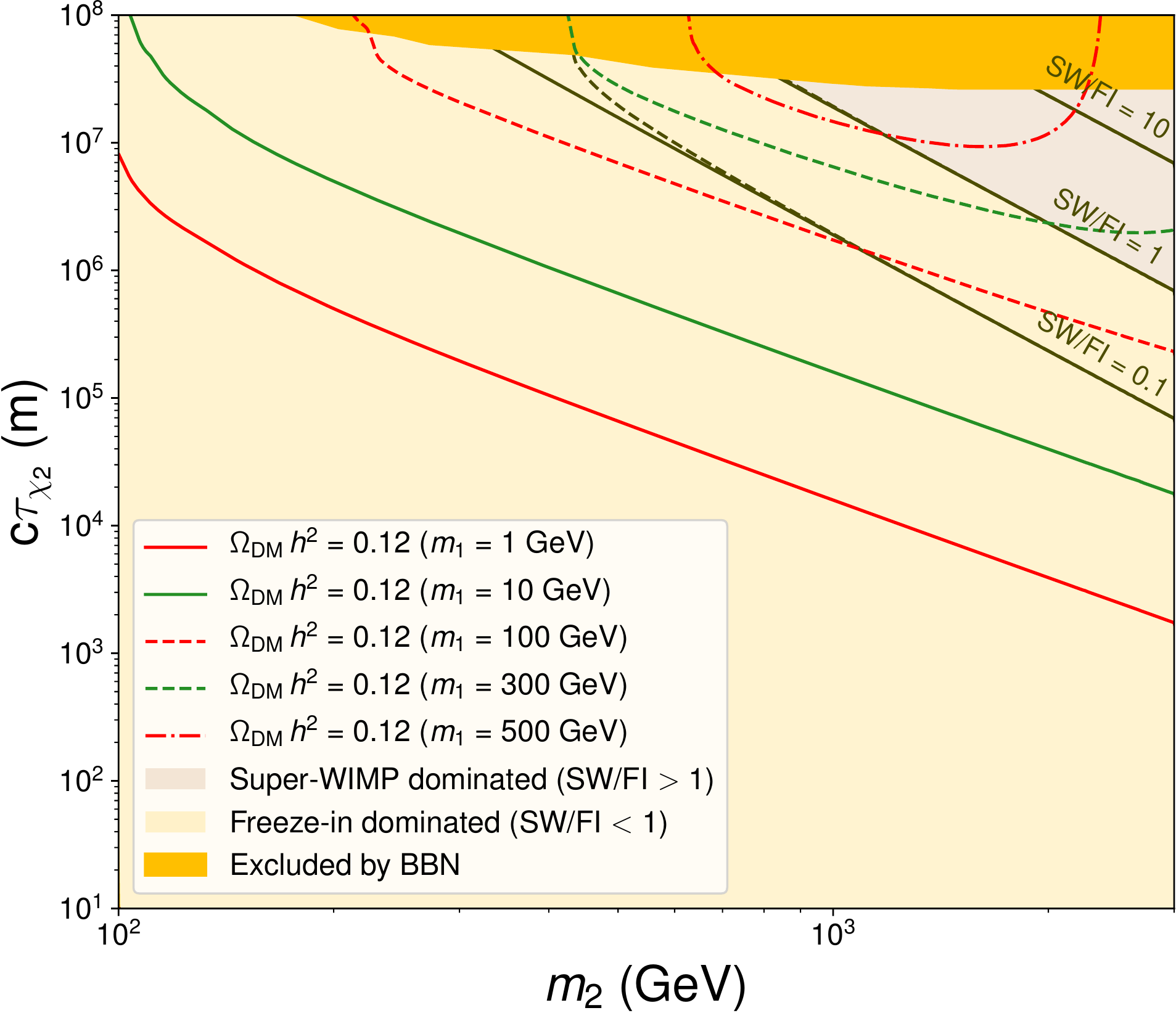}

\vspace{-2mm}
\caption{\small Values of the SW/FI ratio $(m_1\, \Omega_{\mathrm{2}}\,h^2)/(m_2\,\Omega_{\mathrm{1}}\,h^2)$ in 
the ($m_2$, c$\tau_{\chi_2}$) plane for $m_1 = 1$ GeV (solid black lines) and $m_1 = 300$ GeV (dashed black lines). The 
red/green lines correspond to the DM relic abundance condition $\Omega_{\mathrm{DM}}\,h^2 = 0.12$ for various choices of the DM mass $m_1$. The Yellow region is excluded by BBN (see Section~\ref{cosmology_Sec})}
\label{superWIMP}
\end{center}
\end{figure}

For decay widths of the parent particle significantly smaller than its inverse freeze-out timescale
\be
\label{Gamma_freeze_out_timescale}
\Gamma_A \ll \frac{(m_A/20)^2}{\sqrt{(90/(32\, \pi^3 g_{*}))}\, M_{\mathrm{Pl}}} \sim 0.085 \, \left(\frac{m_A}{M_{\mathrm{Pl}}} \right) \times m_A \, ,
\ee
which in the present scenario roughly corresponds to $c \tau_{\chi_2} \gg (100 \,\mathrm{GeV}/m_2)^2 \times 3$ meters, the thermal freeze-out of $\chi_2$ followed by its decay into $\chi_1$ will also contribute to the DM relic density. The contribution to the DM relic abundance from this mechanism is given by $\Omega_2\,h^2 \times m_1/m_2$, where $\Omega_2\,h^2$ is the freeze-out abundance of $\chi_2$. Assuming that $\chi_2$ undergoes freeze-out before decaying as implied by Eq.~\eqref{Gamma_freeze_out_timescale}, the freeze-out abundance of $\chi_2$ corresponds to that of 
thermal Higgsino DM, which for $m_2 \gg m_W$ is simply given by~\cite{ArkaniHamed:2006mb}
\be
\label{omega_2}
\Omega_{\mathrm{2}}\,h^2 \simeq 0.1 \left(\frac{m_2}{\mathrm{TeV}}\right)^2 \, .
\ee
When $\Omega_2\,h^2 \times m_1/m_2$ yields the dominant contribution to the 
DM relic density, it is usually referred to as the super-WIMP~\cite{Feng:2003xh} mechanism (see also~\cite{Garny:2018ali} for a recent work on the interplay between the freeze-in and super-WIMP scenarios). 
Combining the super-WIMP and freeze-in contributions to the DM relic density, we get the total DM relic abundance
\begin{eqnarray}
\label{DM_Relic_Higgs_Total}
\Omega_{\mathrm{DM}}\,h^2 &=& \Omega_{\mathrm{1}}\,h^2 + \Omega_{\mathrm{2}}\,h^2 \times \frac{m_1}{m_2} \, .
\end{eqnarray}
The above discussion allows to quantify the relative importance of the super-WIMP and freeze-in contributions to the DM relic abundance, with their ratio (SW/FI) given by 
$(m_1\, \Omega_{\mathrm{2}}\,h^2)/(m_2\,\Omega_{\mathrm{1}}\,h^2)$. 
In Figure~\ref{superWIMP} we show the SW/FI ratio in the ($m_2$, c$\tau_{\chi_2}$) plane, 
fixing here for simplicity $T_{\mathrm{EW}} = 160$ GeV and two benchmark values for the DM mass, $m_1 = 1$ GeV and $m_1 = 300$ GeV 
(we nevertheless note that the SW/FI ratio is approximately independent of $m_1$, except for $m_2 - m_1 \to m_h$, as can be seen in Figure~\ref{superWIMP}). 
We also show in Figure~\ref{superWIMP} the corresponding DM relic abundance condition $\Omega_{\mathrm{DM}}\,h^2 = 0.12$ using {\tt micrOMEGAs5.0} and 
Eq.~\eqref{DM_Relic_Higgs_Total} for various choices of the DM mass $m_1$. 

As is apparent from Eq.~\eqref{omega_2}, in the present scenario the super-WIMP mechanism could only account for a significant fraction of the observed DM relic density for $m_2 > 1.1$ TeV. In addition large 
DM masses $m_{1} \gtrsim 300$ GeV are required, as shown in Figure~\ref{superWIMP} and expected in order to partially overcome the otherwise large suppression factor $m_1/m_2$ in the  super-WIMP 
contribution from~\eqref{DM_Relic_Higgs_Total}. Such large DM masses imply large values of $c\tau_{\chi_2}$ for DM not to be overproduced by the freeze-in mechanism, which in turn may become constrained from Big Bang Nucleosynthesis (BBN). We discuss these constraints in the next section.

\section{Constraints on Dark Matter freeze-in from Cosmology}
\label{cosmology_Sec}  

\begin{figure}[t]
\begin{center}
\includegraphics[width=0.475\textwidth]{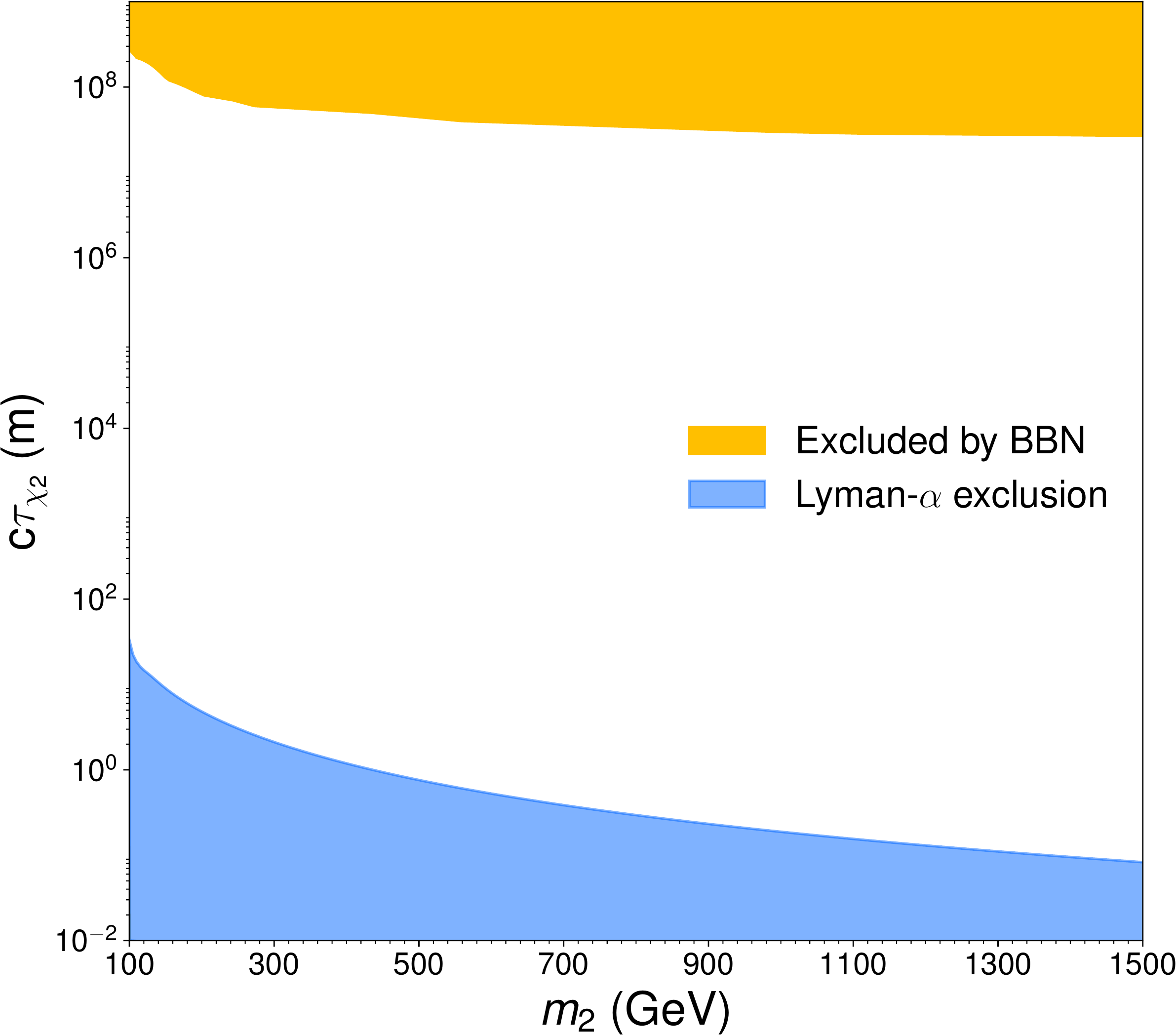}
\hspace{1mm}
\includegraphics[width=0.475\textwidth]{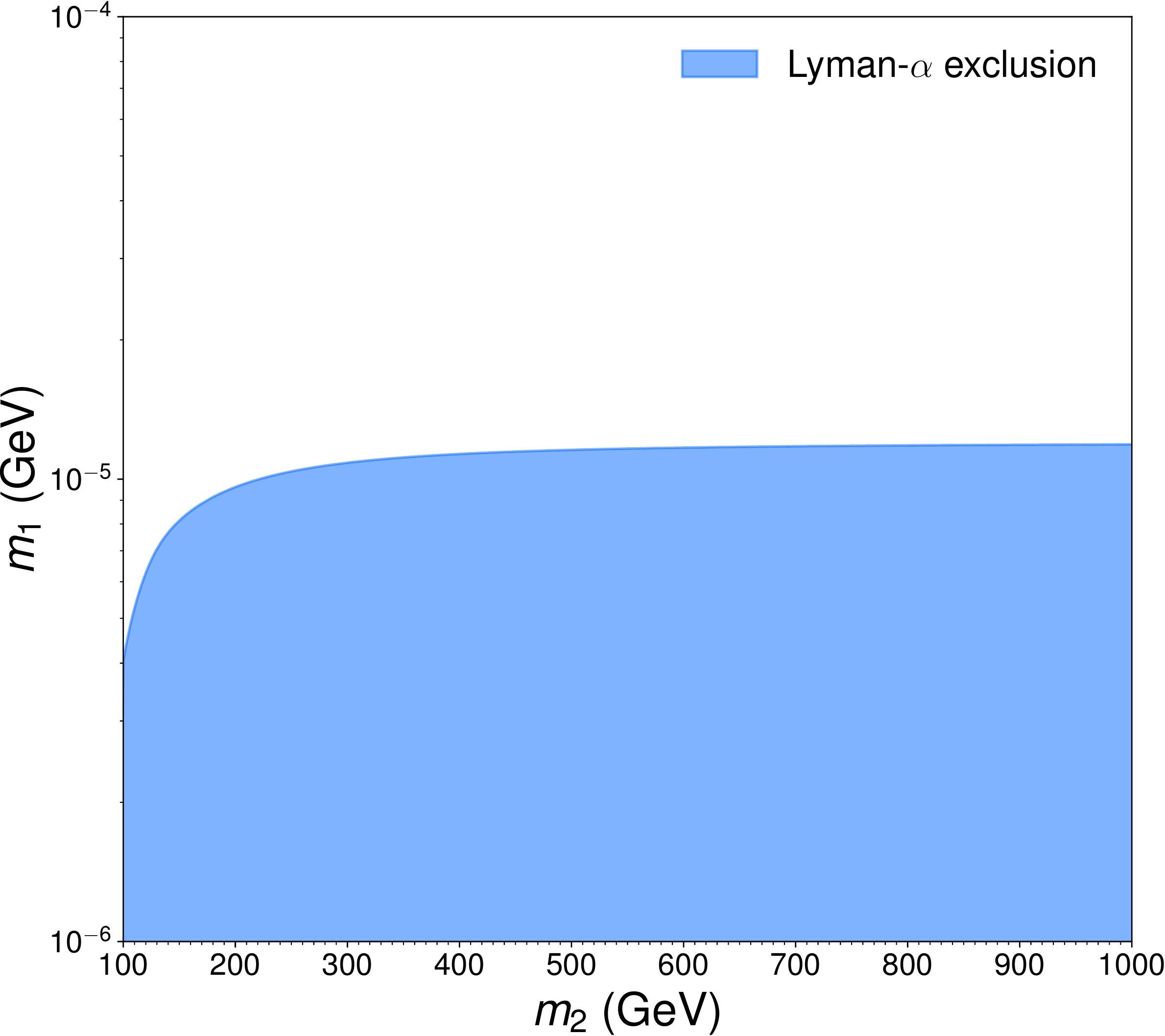}

\vspace{-2mm}
\caption{\small LEFT: Region in the ($m_{2}, c\tau_{\chi_2}$) plane excluded from Big Bang Nucleosynthesis (yellow) and from 
Lyman-$\alpha$ forest observations (blue). 
RIGHT: Lyman-$\alpha$ bound on $m_{1}$ (blue) from Eq.~\eqref{Lalpha_bound} as a function of $m_2$.}
\label{fig:LAlpha_BBN}
\end{center}

\end{figure}

As previously mentioned, there are several important cosmological constraints on this class of freeze-in DM scenarios 
(for a discussion of such constraints in similar scenarios, see also~\cite{Calibbi:2018fqf,Belanger:2018sti}).  
The first constraint we need to consider comes from Big Bang Nucleosynthesis (BBN), which accurately explains the measured primordial 
abundances of light elements in the Universe\footnote{With the only potential exception of $^7$Li, see e.g.~\cite{Goudelis:2015wpa} for a discussion.} 
(see e.g.~\cite{Iocco:2008va,Cyburt:2015mya}). If the state $\chi_2$ lives long enough to decay during BBN, its visible decay products may 
induce several processes that alter the predictions of 
BBN\footnote{For even longer lifetimes $\tau \gtrsim 10^6$ seconds, constraints from the Cosmic Microwave Background (CMB) spectral distorsion become very strong (see~\cite{Hu:1993gc,Kawasaki:2017bqm}). Such long lifetimes are however not relevant for our study.}
In order to obtain the corresponding bounds on the parameter space 
of $m_1$, $m_2$ and c$\tau_{\chi_2}$, we use the recent re-analysis of BBN constraints on long-lived decaying particles 
from~\cite{Kawasaki:2017bqm}. Concerning the hadronic final states $\bar{q}q$ (with $q$ either a first/second generation quark or a $b$ quark), 
ref.~\cite{Kawasaki:2017bqm} derives bounds on the amount of energy injected in the radiation bath in the form of hadrons from 
the decay of a long-lived particle $X$, as a function of the particle lifetime $\tau_X$ and under the assumption that $X$ decays dominantly 
into $\bar{q}q$\footnote{The corresponding limits for the $\bar{b}b$ and $\bar{u}u$ final states found in~\cite{Kawasaki:2017bqm} are very similar, 
and thus approximately valid for scenarios where both decays into $b$-quark pairs and light-quark pairs are possible.}, meaning 
that the hadronic energy injection per particle decay is $E_{\bar{q}q} \sim m_X$. In the present case the hadronic 
energy injection from the decay of $\chi_2$ corresponds only to the energy fraction carried by the visible decay products of $\chi_2$, either 
$h$ or $Z$ (both decays $\chi_2 \to h \chi_1$ and $\chi_2 \to Z \chi_1$ are present during the BBN epoch, which occurs for temperatures 
much below the EW phase transition temperature $T_{\mathrm{EW}}$). This is given by
\bea
E_{\bar{q}q} \simeq \sqrt{m_h^2 + \frac{m_{2}^2}{4} \left[ \left(1 - \xi_h -\xi_1 \right)^2 - 4 \xi_h \xi_1    \right] }  \quad \quad (\chi_2 \to h \chi_1) \nonumber \\
E_{\bar{q}q} \simeq \sqrt{m_Z^2 + \frac{m_{2}^2}{4} \left[ \left(1 - \xi_z -\xi_1 \right)^2 - 4 \xi_z \xi_1    \right]   } \quad  \quad (\chi_2 \to Z \chi_1) 
\eea
with $\xi_h = m_h^2/m_2^2$, $\xi_z = m_Z^2/m_2^2$ and $\xi_1 = m_1^2/m_2^2$.
From~\cite{Kawasaki:2017bqm} we then obtain the limits on the total hadronic 
energy injection onto the thermal bath from $\chi_2$ decays, given by
\be
\frac{\Omega_{\mathrm{2}}\,h^2}{m_2} \times \sum_{a = h,Z}  E^{a}_{\bar{q}q} \times \mathrm{BR}(\chi_2 \to a\, \chi_1)  \, \,\mathrm{BR}(a \to \mathrm{hadrons}) 
\ee
with $\Omega_{\mathrm{2}}\,h^2$ from~\eqref{omega_2}. Assuming $m_1 \ll m_2, m_h$, the corresponding limits on the ($m_2$, c$\tau_{\chi_2}$) parameter space are shown in Figure~\ref{fig:LAlpha_BBN} (left). We note that 
leptonic decays of $\chi_2$ also affect BBN, yet these constraints are found to be significantly weaker than the hadronic ones (see~\cite{Kawasaki:2017bqm}) and are therefore not relevant in the present case.

\vspace{2mm}

A second important set of limits originate from constraints on the washout of small-scale structure by DM with a non-negligible 
velocity dispersion (partially relativistic, or {\sl warm}). The leading such constraint comes from Lyman-$\alpha$ forest observations. 
For thermal warm dark matter (WDM) these observations place a lower limit on the DM mass in the range $m_{\rm DM} \gtrsim 4.09 - 5.3$ keV~\cite{Baur:2015jsy,Yeche:2017upn,Irsic:2017ixq}. 
In two recent works~\cite{Heeck:2017xbu,Boulebnane:2017fxw} the 
Lyman-$\alpha$ limit for the case of freeze-in DM produced via two-body decays of a parent particle in thermal equilibrium with the plasma (which is precisely our scenario) has been estimated by comparing the suppression in the linear matter power spectrum from a thermal WDM scenario 
(with $m_{\rm DM}$ given by the WDM Lyman-$\alpha$ limit, taken to be 
$4.65$ keV~\cite{Yeche:2017upn,Baur:2017stq}) to that of
the freeze-in scenario. The Lyman-$\alpha$ bound in this case reads
\begin{equation}
m_{\rm DM} \gtrsim 12 \ {\rm keV} \left( \frac{ \sum_{ij} g_i \Gamma_{ij}  \Delta_{ij}^\eta}{\sum_{ij} g_i \Gamma_{ij} }\right)^{1/\eta} \, ,
\label{Lalpha_bound}
\end{equation}
where the sum runs over all decay channels of the parent particle(s) that contribute to DM production, $\Gamma_{ij}$ are the corresponding decay widths
$\Gamma(A_i \to B_j \chi_1)$ (in our case given by Eq.~\eqref{decayFIDM_1}), $g_i$ are the number of degrees of freedom of the parent particle $A_i$,  
the parameter $\Delta_{ij}$ yields the mass splitting between the parent particle and the visible decay product in each channel 
(e.g.~$\Delta_{ij} = 1 - m_{Z}^2/m_{2}^2$ for the decay $\chi_2 \to Z \chi_1$) and $\eta \simeq 1.9$~\cite{Boulebnane:2017fxw}. 
Assuming that freeze-in DM saturates the observed DM relic density,
Figure~\ref{fig:LAlpha_BBN} (right) shows the Lyman-$\alpha$ bound on $m_{1}$ in the present scenario. This bound can in turn be translated into a lower limit 
for $c\tau_{\chi_2}$, shown in Figure~\ref{fig:LAlpha_BBN} (left).

The combination of BBN and Lyman-$\alpha$ bounds yields an allowed range of parent particle decay lengths $c\tau_{\chi_2}$ for viable DM freeze-in, 
$c\tau_{\chi_2} \sim 10^{-1} - 10^7$ meters, which provide a clear target for LLP searches at colliders. We investigate the current and projected sensitivity of such searches to the Higgs-assisted freeze-in from decay DM scenario analyzed in this work in sections~\ref{MATHUSLA_Sec} - \ref{100TeV_Sec}.

\section{Probing freeze-in Dark Matter with the MATHUSLA detector}
\label{MATHUSLA_Sec}

Considering the limit $m_{2} \gg m_1,\,m_W,\,m_Z,\,m_h$ in combination with the freeze-in DM relic abundance condition $\Omega_{\mathrm{DM}} h = 0.12$ yields an estimate for the decay length of $\chi_2$ (see also Eq.~\eqref{lifetime_A})
\begin{equation}
\label{Lifetime_FIHiggs}
c \tau_{\chi_2} \sim 4 \,\,\text{Km }\times \left(\frac{m_1}{100 \, \MeV}\right) \left(\frac{500 \, \GeV}{m_2}\right)^2\,\,.
\end{equation}
Such long lifetimes make it challenging to search for $\chi_2$ with the LHC's ATLAS and CMS detectors (we will nevertheless discuss the current 
ATLAS and CMS bounds, as well as future prospects in Section~\ref{Sec:ATLAS_CMS}),
except maybe for rather small values of $m_1$ close to the Lyman-$\alpha$ bound~\cite{Calibbi:2018fqf}.
Here we analyze the prospects for probing the parameter space of the present scenario with the proposed MATHUSLA surface 
detector for long-lived particles~\cite{Chou:2016lxi,Curtin:2017izq,Curtin:2018mvb,Alpigiani:2018fgd}.
The large detector volume of MATHUSLA together with its capability to 
effectively operate as a background-free environment for LLP searches (see~\cite{Curtin:2018mvb,Alpigiani:2018fgd} for details) at the LHC, yields an ideal setup to probe freeze-in DM scenarios.

At the LHC, the states $\chi_2$ and $\psi^{\pm}$ can be produced via the Drell-Yan processes $p p \to \chi_2 \chi_2$, 
$p p \to \chi_2 \psi^{\pm}$, $p p \to \psi^{+}\psi^{-}$. For our study, we implement the Lagrangian from Eq.~\eqref{eq:DMSM_FI} in {\tt FeynRules}~\cite{Alloul:2013bka} and simulate the various Drell-Yan production processes for $\chi_2$ and $\psi^{\pm}$ at $\sqrt{s} = 13$ TeV LHC in {\tt Madgraph$\_$aMC@NLO}~\cite{Alwall:2014hca}. We then normalize the respective cross sections to the corresponding $\sqrt{s} = 13$ TeV LHC NLO+NLL charged/neutral Higgsino production cross sections $\sigma^{\mathrm{LHC}}_{13}$ computed with 
{\tt Resummino-2.0.1}~\cite{Fuks:2012qx,Fuks:2013vua}\footnote{These are given also by the CERN LHC SUSY XS Working Group~\cite{SUSYworkinggroup}.}
(with the PDF set {\sc MSTW2008nlo90cl} from {\tt LHAPDF}~\cite{Buckley:2014ana}). The Higgsino pair production cross sections for the various Drell-Yan processes are shown explicitly in Appendix~\ref{AppendixC}. In the following, we also consider that the decays $\psi^{\pm} \to \chi_2 + X$ ($X = \pi^{\pm},\,\ell^{\pm} \nu$ being very soft), which occur within a few cm of the interaction point, effectively convert all the charged states $\psi^{\pm}$ into neutral $\chi_2$ ones with approximately identical kinematics.

The MATHUSLA detector would be located at a distance $\sim \mathcal{O}(100\,\mathrm{m})$ from the ATLAS or CMS interaction point. 
The probability for a $\chi_2$ particle to decay inside the MATHUSLA detector volume is given by 
\begin{equation}
\label{Prob_FI_Mathusla_chi2}
P_\mathrm{decay}^\mathrm{MATH} = \epsilon_{\mathrm{geometric}} \times 
P_\mathrm{decay}(\beta\, c \tau_{\chi_2}, L_a, L_b)
\end{equation}
with $P_\mathrm{decay}(\beta \, c \tau_{\chi_2}, L_a, L_b)$ given by 
\begin{equation}
\label{e.Pdecay}
P_\mathrm{decay}(\beta\, c \tau_{\chi_2}, L_a, L_b) = e^{- \frac{L_a}{\beta\, c \tau_{\chi_2}}} - e^{- \frac{L_b}{\beta\, c \tau_{\chi_2}}}\,\,.
\end{equation}
Here $L_a$ and $L_b$ are the distances from the interaction point for which $\chi_2$ respectively enters and leaves the MATHUSLA detector volume,  
and $\beta$ is the $\chi_2$ boost factor $|\vec{p}_{\chi_2}|/m_2$. 
In this work we consider two possible MATHUSLA design configurations: MATHUSLA100 and MATHUSLA200. In the former case, the MATHUSLA detector volume for LLP decays as measured from the interaction point is~\cite{Alpigiani:2018fgd} $x \in [-50, \,50]$ m, $y \in [100, \,120]$ m, $z \in [100, \,200]$ m, while in the latter case (corresponding to the original MATHUSLA proposal~\cite{Chou:2016lxi,Alpigiani:2018fgd}) the detector volume 
as measured from the interaction point 
is $x \in [-100, \,100]$ m, $y \in [100, \,120]$ m, $z \in [100, \,300]$ m.
The MATHUSLA geometric acceptance 
$\epsilon_{\mathrm{geometric}}$ in Eq.~\eqref{Prob_FI_Mathusla_chi2} then measures the fraction of generated events 
contained within the MATHUSLA solid angle as seen from the 
interaction point, which we extract from our 
event simulation\footnote{For an isotropic distribution of $\chi_2 \bar{\chi}_2$ events, $\epsilon_{\mathrm{geometric}}$
would correspond to twice the MATHUSLA detector solid angle from the interaction point,
$\epsilon_{\mathrm{geometric}} \simeq 2\times \frac{1}{4\pi} \int_{-\pi/4}^{\pi/4} d \phi \int^{\pi/4}_{{\rm arctan}(12/30)} {\rm sin} \theta\, d \theta \sim 0.05$ for MATHUSLA200 and $\epsilon_{\mathrm{geometric}} \simeq 2\times \frac{1}{4\pi} \int_{-{\rm arctan}(1/2)}^{{\rm arctan}(1/2)} d \phi \int^{\pi/4}_{{\rm arctan}(12/20)} {\rm sin} \theta\, d \theta \sim 0.022$ for MATHUSLA100.
In a more general case the convolution with the $\eta$, $\phi$ dependence of the event sample modifies this estimate.}.
Specifically, for the computation of $\epsilon_{\mathrm{geometric}}$ we consider events for which the $\chi_2$ trajectory intersects the MATHUSLA 
tracking layers (see~\cite{Chou:2016lxi, Curtin:2017izq} for details on the tracking layout within the MATHUSLA detector) and assume for simplicity
that the visible $\chi_2$ decay products would then also hit the tracking layers if the decay happens inside MATHUSLA (for $m_2 \gg m_h + m_1$ this occurs automatically, since the decay products of $\chi_2$ will be fairly collinear to the $\chi_2$ trajectory, while for $m_2 \to m_h + m_1$ addressing the possible modification of our acceptances requires a more detailed event and detector simulation beyond the scope of this work).
We also assume perfect MATHUSLA detection efficiency for $\chi_2$ decays inside the detector volume. 

\begin{figure}[h]

\centering

$\vcenter{\hbox{\includegraphics[width=0.7\textwidth]{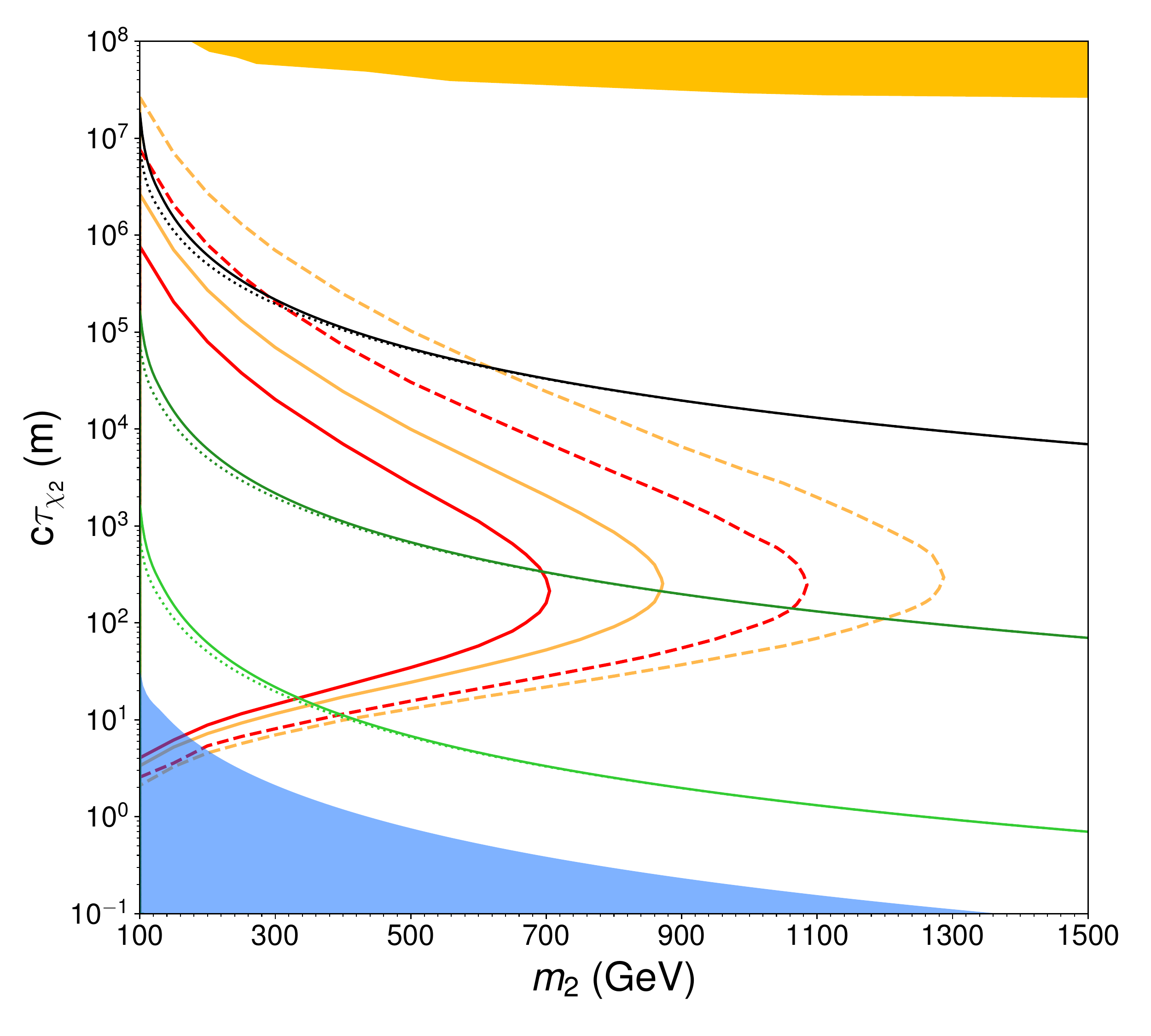}}}$
$\vcenter{\hbox{\includegraphics[width=0.29\textwidth]{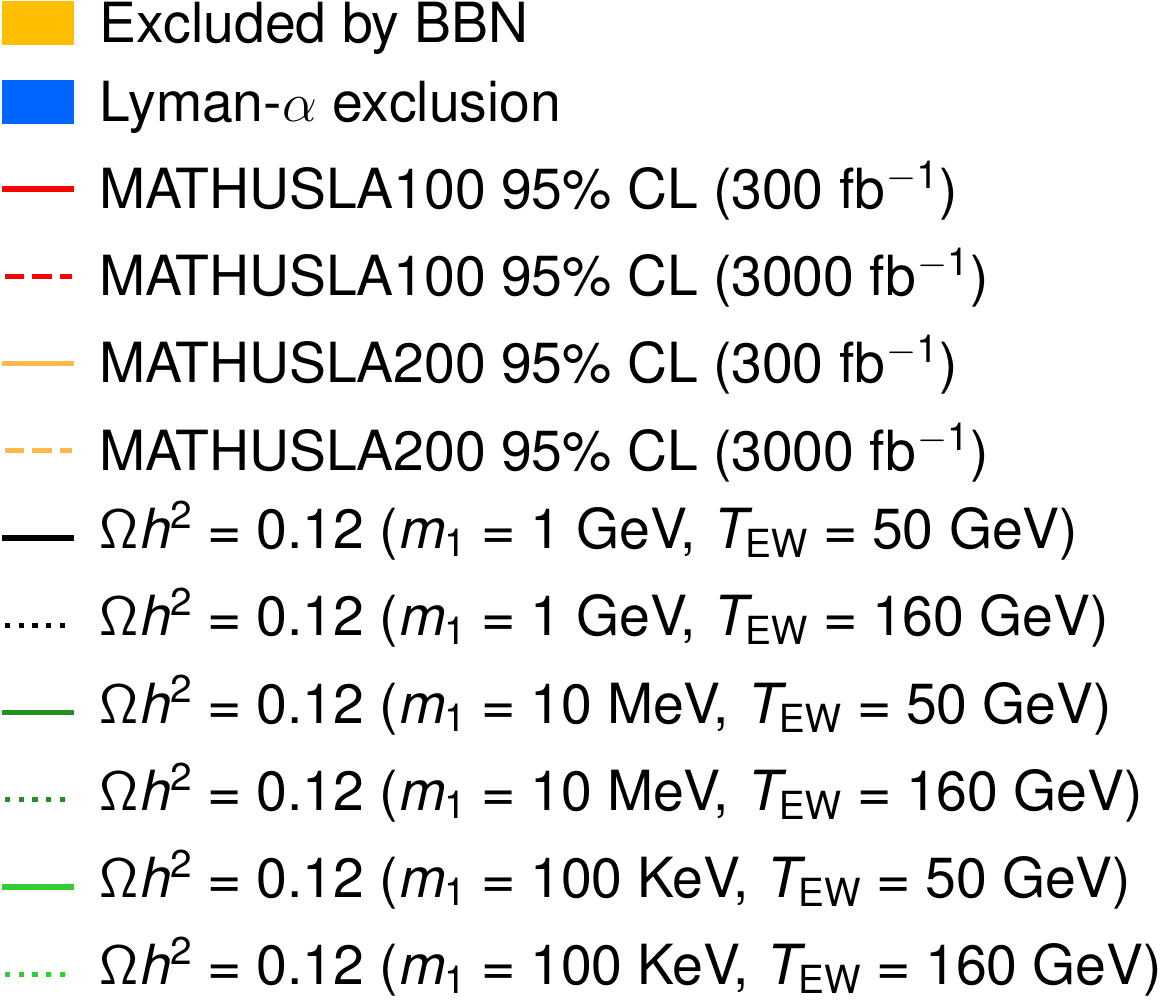}}}$

\caption{\small 95\% C.L. exclusion sensitivity for MATHUSLA200 (orange) and 
MATHUSLA100 (red) in the ($m_2$, $c\tau_{\chi_2}$) plane, for 300 fb$^{-1}$ (solid lines) and 3000 fb$^{-1}$ (dashed lines). Lines yielding 
the observed DM relic density $\Omega_{\mathrm{DM}} h^2 = 0.12$ (obtained with {\tt micrOMEGAs5.0}) 
are shown for $m_1 = 1$ GeV (black), $10$ MeV (dark green), $100$ KeV (light green), together with the bounds from 
BBN (yellow) and Lyman-$\alpha$ (blue).
}
\label{FI_Higgs_m1m2}

\end{figure}

The number of expected signal events at MATHUSLA is then given by
\begin{equation}
N_{\mathrm{events}} = \sigma^{\mathrm{LHC}}_{13} \times \mathcal{L} \times \int P_\mathrm{decay}^\mathrm{MATH} \, ,
\end{equation}
where $\mathcal{L}$ is the integrated luminosity and the integral over $P_\mathrm{decay}^\mathrm{MATH}$ denotes a generic integration over phase-space and decay length for the signal event sample.
Considering the MATHUSLA detector as an essentially background-free environment (see~\cite{Alpigiani:2018fgd} for a detailed discussion on this point), the 95\% C.L. exclusion sensitivity corresponds to $N_{\mathrm{events}} = 3$.

In Figure~\ref{FI_Higgs_m1m2} we show the MATHUSLA100 and MATHUSLA200 95\% C.L. sensitivities in the ($m_2$, $c\tau_{\chi_2}$) plane for 
an integrated luminosity $\mathcal{L} = 300$ fb$^{-1}$ (solid) and $\mathcal{L} = 3000$ fb$^{-1}$ (dashed).
We also show the lines yielding the observed DM relic abundance for $m_1 = 1$ GeV (black), $m_1 = 10$ MeV (dark green) and $m_1 = 100$ KeV (light green) obtained with {\tt micrOMEGAs5.0}, assuming the temperature of the 
EW phase transition to be in the range $T_{\mathrm{EW}} \in [50\,\mathrm{GeV}, \, 160\, \mathrm{GeV}]$.
The results of Figure~\ref{FI_Higgs_m1m2} show that MATHUSLA could cover a wide region of the viable freeze-in parameter space
between the Lyman-$\alpha$ and BBN constraints, and probe DM masses up to $m_1 \sim 1 - 10$ GeV. 
In the next section, we compare these sensitivity projections with those of searches by ATLAS and CMS, to investigate the complementarity between MATHUSLA
and the existing LHC detectors.

\section{Searches for freeze-in Dark Matter with ATLAS and CMS}
\label{Sec:ATLAS_CMS}

\subsection{Standard Dark Matter searches: $E^{\rm miss}_T$ (mono-$X$) signatures}
\label{sec:monoJet}

\begin{figure}[h]
\begin{center}
\includegraphics[align=c,width=0.3\textwidth]{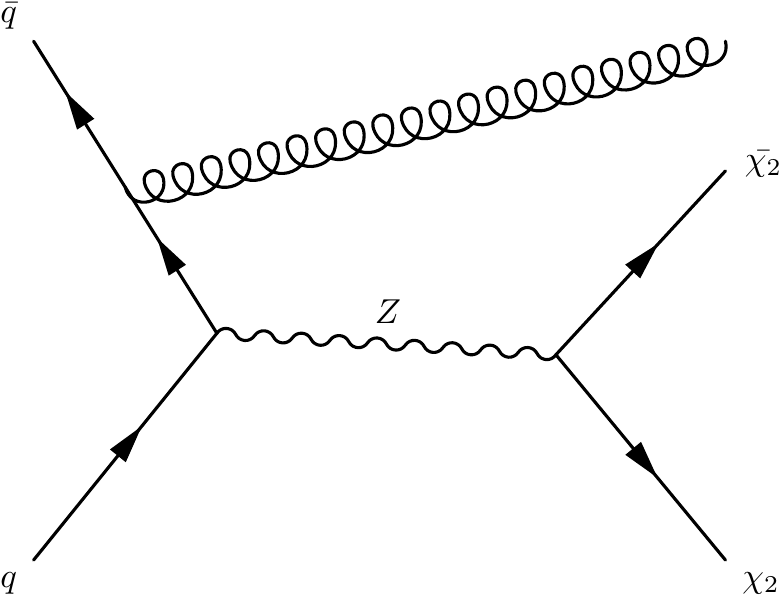}
\hspace{10mm}
\includegraphics[align=c,width=0.6\textwidth]{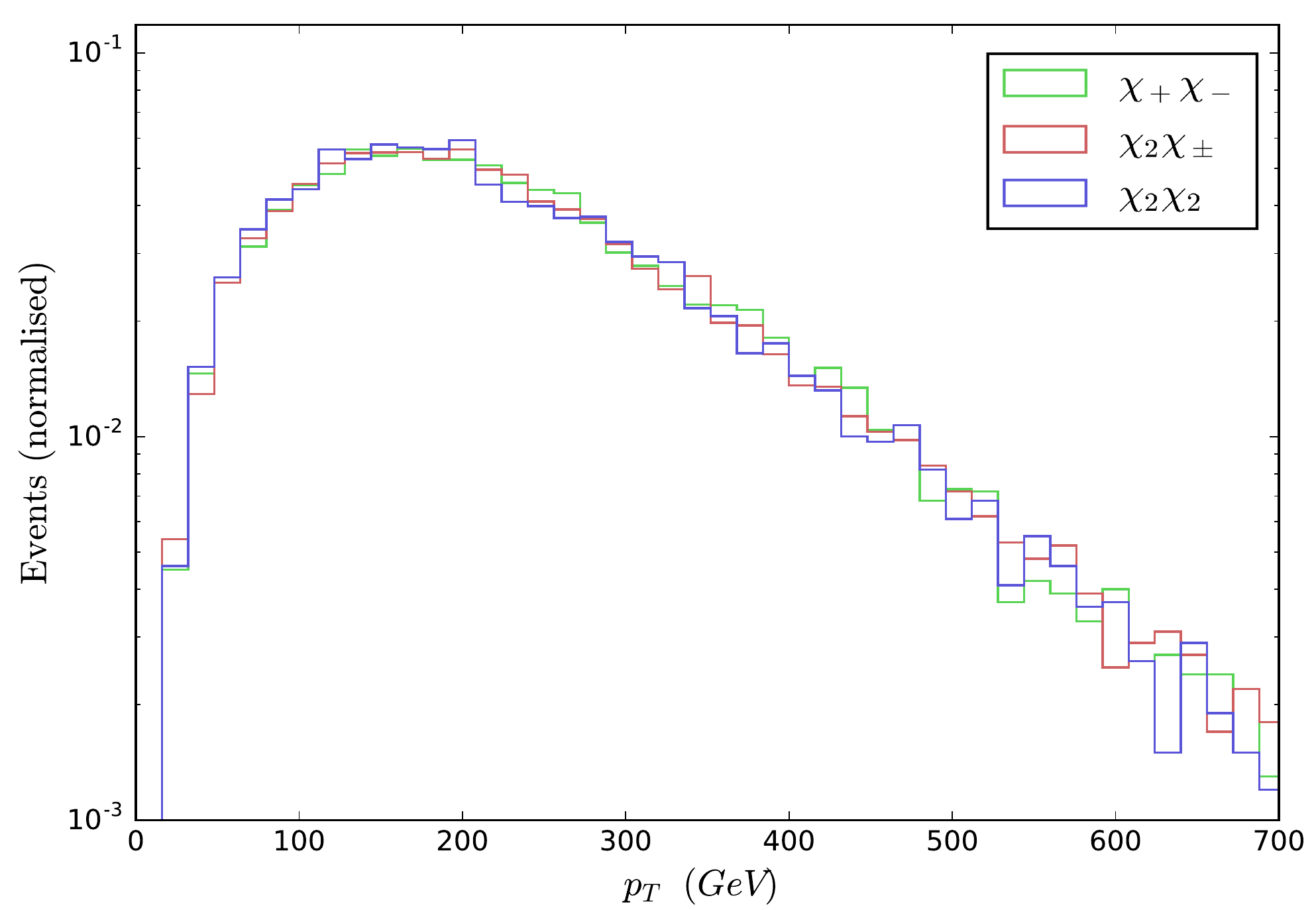}

\caption{\small LEFT: Feynman diagram for the process $p p \to \bar{\chi_2} \chi_2 + j$. RIGHT: 
Jet $p_T$ distribution for $p p \to \bar{\chi_2} \chi_2 + j$ (blue), $p p \to \chi_2 \psi^{+} + j$ (red), 
$p p \to \chi_2 \psi^{-} + j$ (green) and $p p \to \psi^{+}\psi^{-} + j$ (orange) for $m_2 \simeq m_{\psi^{\pm}} = 300$ GeV. 
}
\label{DM_mono_jet}
\end{center}

\vspace{-5mm}

\end{figure}

Due to the long lifetime of $\chi_2$, standard ``mono-$X$" searches for DM at the LHC may be sensitive to the freeze-in parameter space when the 
decay of $\chi_2$ happens outside the ATLAS/CMS detector. Here we analyze the sensitivity of such ``traditional"  DM searches, focusing on the mono-jet signature which generally yields stronger constraints as compared e.g.~to mono-$\gamma$. Considering the state $\chi_2$ to be invisible (decaying outside the relevant detector volume),  jet + $E^{\rm miss}_T$ signatures can be obtained in the current scenario via Drell-Yan production $p p \to \bar{\chi_2} \chi_2$, $p p \to \chi_2 \psi^{\pm}$, $p p \to \psi^{+}\psi^{-}$ in association with an initial-state-radiation (ISR) jet, see Figure~\ref{DM_mono_jet} (left). In the latter two processes $\psi^{\pm}$ decays into $\chi_2$ and a very soft $\pi^{\pm}$. The particle $\psi^{\pm}$ will in principle leave a short track in the detector's inner tracker before decaying, but since the decay length of $\psi^{\pm}$ is $\mathcal{O}(1\,\mathrm{cm})$ (see Eq.~\eqref{Gamma_charged} and Section~\ref{sec:DT}), such a short track typically does not affect the mono-jet event selection\footnote{The mono-jet analysis features only indirect track vetoes, in the form of well-identified electrons, muons and hadronic $\tau$ leptons, all of which require further activity in the calorimeters or muon stations. That way, the processes 
$p p \to \chi_2 \psi^{\pm}$, $p p \to \psi^{+}\psi^{-}$ with an ISR jet may also contribute to the mono-jet signal. We thank Steven Lowette for clarifications on this point.}.   

We first focus on the Drell-Yan $p p \to \bar{\chi_2} \chi_2$ process with an ISR jet, which is mediated by an off-shell $Z$-boson. 
This scenario can be reinterpreted as a simplified model for DM (with $\chi_2$ as the invisible particle)
with a vector mediator~\cite{Abdallah:2015ter,Boveia:2016mrp} with $m_{\rm med} = m_Z$, and with the mediator 
couplings to SM quarks ($g_{q}$) and to $\chi_2$ ($g_{\chi}$) being fixed by 
the respective gauge quantum numbers of the quarks and $\chi_2$. This allows us to use the existing constraints on simplified DM models with vector 
mediators from the CMS mono-jet analysis at $\sqrt{s} = 13$ TeV with an integrated luminosity $\mathcal{L} =  35.9$ fb$^{-1}$~\cite{Sirunyan:2017jix}.
Moreover, since the $p_T$ distribution of the ISR jet is similar for all three 
processes $p p \to \bar{\chi_2} \chi_2$, $p p \to \chi_2 \psi^{\pm}$, $p p \to \psi^{+}\psi^{-}$, as 
shown in Figure~\ref{DM_mono_jet} (right), we can to a good approximation rescale the cross section 
for $p p \to \bar{\chi_2} \chi_2 + j$ by the total cross section including all three processes (noting also 
that there is no interference among the different processes).

\begin{figure}[t]
\begin{center}
\includegraphics[width=0.6\textwidth]{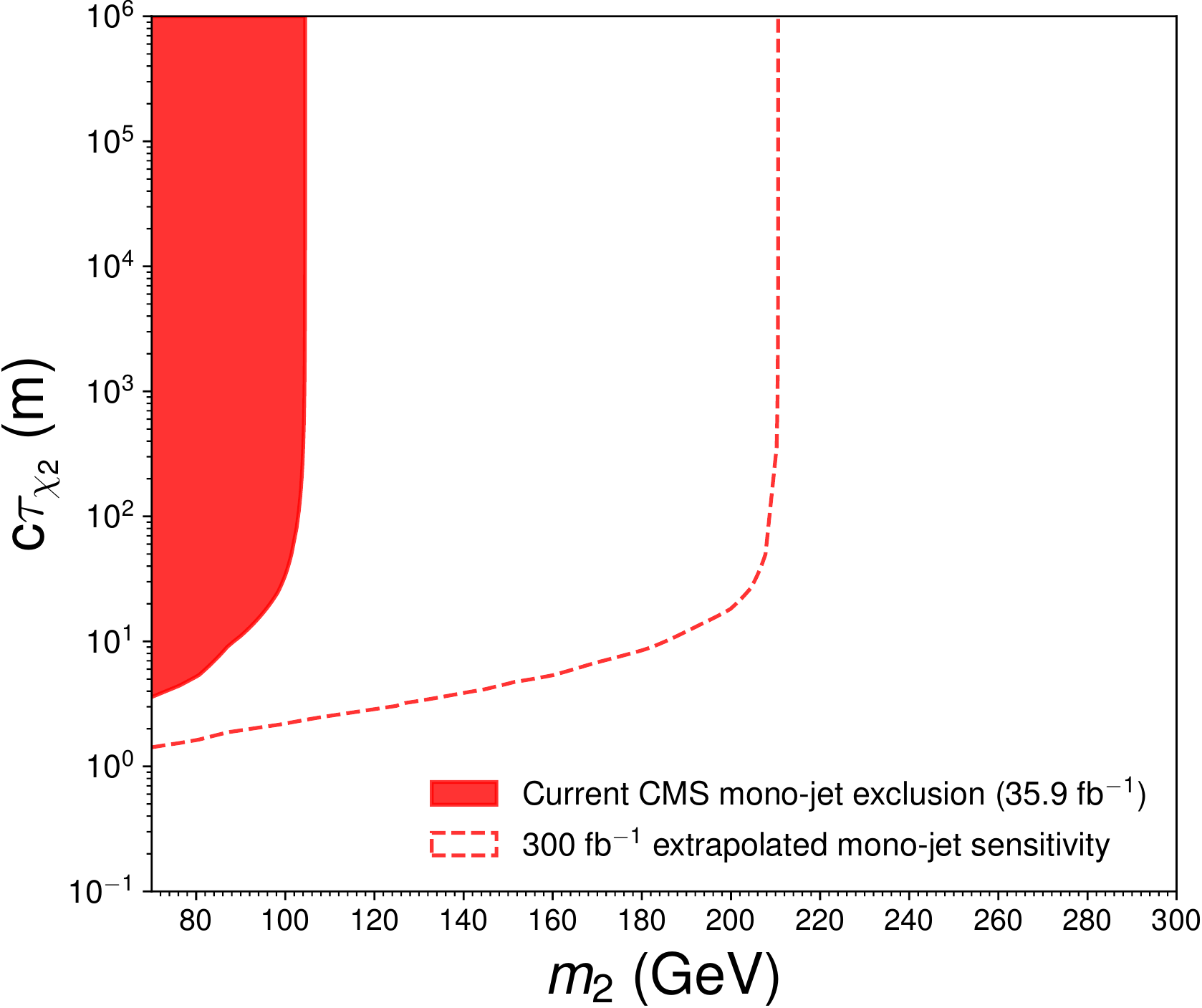}

\caption{\small Mono-jet 95\% C.L. exclusion bound in the ($m_2,\,\mathrm{c}\tau_{\chi_2}$) plane for $35.9$ fb$^{-1}$ (solid) and naive
extrapolation to 300 fb$^{-1}$ (dashed), see text for details.}
\label{DM_mono_jet_2}
\end{center}

\end{figure}

We then use the publicly provided 95\% C.L. exclusion limit on the mono-jet signal strength $\mu_{\mathrm{CMS}}$ 
(mono-jet cross section divided by the theoretical cross section for a DM simplified model with a vector mediator and $g_q = 0.25$, $g_{\chi} = 1$) 
as a function of the DM mass from the CMS analysis~\cite{Sirunyan:2017jix} for $m_{\rm med} = m_Z$, weighted by the decay probability of 
$\chi_2$ inside the CMS detector. This probability is approximately given 
by ${\rm exp}[-\bar{L}_{{\rm CMS}}/c \tau_{\chi_2}]$, with $\bar{L}_{{\rm CMS}}$ a mean CMS relevant 
detector size taken here to be the average hadronic calorimeter radius, $\bar{L}_{{\rm CMS}} \sim 2.5$ 
meters\footnote{More accurately, the detector geometry should be convoluted with the rapidity distribution 
of our signal events and the probability of vetoing a signal event in the mono-jet analysis if the decay happens 
in different parts of the CMS detector. This is however beyond the scope of this work. The exponential fall-off of 
the mono-jet sensitivity for small values of c$\tau_{\chi_2}$ shown in Figure~\ref{DM_mono_jet_2} should then only be taken as approximate.}.
The corresponding CMS 95\% C.L. exclusion limit on the present scenario is then given by equating
\begin{equation}
\label{eqCMSmonojet}
\mu_{{\rm CMS}} (m_2) = \frac{g^2}{4\,c_{\theta_W}^2} \, \frac{(g^Z_L)^2 + (g^Z_R)^2}{(0.25)^2} \times r_{\sigma} (m_2)
\times {\rm exp}[-\bar{L}_{{\rm CMS}}/c \tau_{\chi_2}]
\end{equation}
where $\mu_{\mathrm{CMS}}(m_2)$ is the CMS exclusion limit given in~\cite{Sirunyan:2017jix} for $m_{\mathrm{DM}} = m_2$ and $m_{\rm med} = m_Z$, 
$g^Z_L$ and $g^Z_R$ are the couplings of the $Z$-boson to the left and right-handed SM quarks\footnote{Here we take the up-type quark 
couplings, which yield the dominant contribution, to compute the exclusion limit. In practice, both the up and down-type quark contributions should be 
taken into account, bearing in mind that they are slightly different and this difference can be taken into account by re-weighting the 
different PDF contributions to the Drell-Yan process as a function of the momentum transfer. We however do not perform such re-weighting here.} 
and $r_{\sigma} (m_2)$ is the ratio between the cross section for $p p \to \bar{\chi_2} \chi_2 + j$ and the total cross section including 
all other Drell-Yan processes (which also involve $\psi^{\pm}$ production), which we compute using {\tt Madgraph$\_$aMC@NLO} at leading order. 
From Eq.~\eqref{eqCMSmonojet} we derive the current approximate 95\% C.L. bound $m_2 \gtrsim 104$ GeV in the limit $c \tau_{\chi_2}  \gg \bar{L}_{{\rm CMS}}$, see Figure~\ref{DM_mono_jet_2}. 
Using a simple $\sqrt{\mathcal{L}}$ luminosity rescaling, we also provide a naive projection of the limit to $\mathcal{L} = 300$ fb$^{-1}$ 
with $\sqrt{s} = 13$ TeV, shown in Figure~\ref{DM_mono_jet_2}, yielding $m_2 \gtrsim 210$ GeV in the limit $c \tau_{\chi_2}  \gg \bar{L}_{{\rm CMS}}$.
We however stress that since this projection has been derived assuming the SM background uncertainties in the mono-jet search are dominantly statistical, 
and in reality the mono-jet search is systematics dominated and so the true mono-jet sensitivity for $\mathcal{L} = 300$ fb$^{-1}$ should be significantly weaker than the above value of $m_2$. On the other hand, future improvements in the systematic errors would make our (overly aggressive) limit more realistic.

\subsection{Disappearing track signatures}
\label{sec:DT}

Due to the very small mass splitting $\delta m \sim 300$ MeV  between the states $\psi^{\pm}$ and $\chi_2$ (recall Eq.~\eqref{deltam_charged}), 
the state $\psi^{\pm}$ is relatively long-lived, 
with its decay length in the range c$\tau_{\psi^{\pm}} \in [0.7, \,2.1]$ cm for $m_{\psi^{\pm}}$ between 90 GeV and 1 TeV. 
The dominant decay $\psi \to \chi_2\, \pi^{\pm}$ can then lead to a disappearing track signature, due to the softness of the produced pion. 
Both CMS~\cite{Sirunyan:2018ldc} 
and ATLAS~\cite{Aaboud:2017mpt} have performed searches for disappearing tracks with LHC 13 TeV data and an integrated luminosity of 36.1 fb$^{-1}$ 
(38.4 fb$^{-1}$) in the case of ATLAS (CMS). We note that ATLAS can reconstruct tracks as short as $\sim 12$ cm, while for the CMS tracker the minimum 
reconstructed track length is $\sim 25- 30$ cm (see e.g.~the discussion in~\cite{Evans:2016zau, Belanger:2018sti}). As a result, the ATLAS search can constrain smaller 
lifetimes than the CMS study. In fact, the CMS search~\cite{Sirunyan:2018ldc} only constrains chargino lifetimes $\tau > 0.07$ nanoseconds (ns) corresponding to 
$c \tau > 2.1$ cm, already at the edge of the maximum value of $c \tau_{\psi^{\pm}}$ possible in the present scenario. Using the exclusion from the 
$\tau \in [0.07,\,0.1]$ ns, 
$m_2 \in [50,\,150]$ GeV bin in the CMS analysis~\cite{Sirunyan:2018ldc} (yielding $\sigma \times B < 17.60$ pb) and using the chargino production 
cross sections provided in~\cite{SUSYworkinggroup}, we derive the tentative current bound $m_{\psi^{\pm}} > 84$ GeV, which is in fact weaker than 
existing LEP constraints on the state $\psi^{\pm}$~\cite{LEPSUSYWG,Egana-Ugrinovic:2018roi}. 
In contrast, the ATLAS collaboration recently reinterpreted their analysis~\cite{Aaboud:2017mpt} in terms of 
pure Higgsino\footnote{This essentially corresponds to the 
present scenario, the only difference occurring for values of c$\tau_{\chi_2} \lesssim 10$ cm, for which the decay $\psi^{\pm} \to W^{\pm} \chi_1$ 
starts being important (see Figure~\ref{BR_chi2} (right)), which would make the state $\psi^{\pm}$ shorter-lived and thus 
result in a weakening of the 95\% C.L. exclusion limit from disappearing track searches.} 
production~\cite{ATL-PHYS-PUB-2017-019}, obtaining a present 95\% C.L. exclusion of $m_{\psi^{\pm}} \gtrsim 145$ GeV. 
In addition, the ATLAS collaboration has performed a preliminary study of their future sensitivity to a disappearing track signature 
in the pure Higgsino scenario~\cite{ATL-PHYS-PUB-2018-031}, which would yield a 95\% C.L. sensitivity of $m_{\psi^{\pm}} \sim 260$ GeV at the 
HL-LHC with $\sqrt{s} = 14$ TeV and $\mathcal{L} = 3000$ fb$^{-1}$. We show the corresponding limits in Figure~\ref{Figure_FI_DV_MET}.

\subsection{Displaced vertex (DV) signatures}
\label{sec:DV}

Neutral long-lived particles may be searched for via displaced vertex signatures in ATLAS, CMS and LHCb, 
and a wide variety of analyses already exist for LHC $\sqrt{s} = 7$ , $8$ and $13$ TeV 
(see e.g.~\cite{CMS:2014hka,Aaij:2014nma,Aad:2015uaa,Aad:2015rba,Liu:2015bma,CMS:2016isf,Aaboud:2017iio,Aaboud:2018aqj}). 
Among the various searches which could most effectively probe the scenario explored in this work, we highlight: 

\begin{itemize}

\item CMS/ATLAS searches looking for displaced lepton pairs~\cite{CMS:2014hka}, which would be sensitive to 
decays $\chi_2 \to Z \, \chi_1$ ($Z \to \ell\ell$), and to a lesser extent (due to the much smaller leptonic branching fraction) 
to $\chi_2 \to h\, \chi_1$ ($h \to \ell \nu_{\ell}\, \ell^{'} \nu_{\ell^{'}}$).

\item ATLAS searches 
looking for activity in the muon spectrometer~\cite{Aad:2015uaa,Aaboud:2018aqj}, which could in principle be sensitive to very large decay 
lengths $c\tau_{\chi_2} \gg 1$ m.

\item ATLAS searches for displaced vertices + $E^{\rm miss}_T$~\cite{Aaboud:2017iio}. These would take advantage of the 
large hadronic branching fraction of both $Z$ and $h$, as well as of the missing momentum in the decays $\chi_2 \to Z/h \,\chi_1$.

\end{itemize}

In this work, we concentrate on the latter DV + $E^{\rm miss}_T$ search by ATLAS~\cite{Aaboud:2017iio} at $\sqrt{s} = 13$ 
TeV with $32.8$ fb$^{-1}$, leaving an exploration of the other two searches highlighted above for future work.

The ATLAS DV + $E^{\rm miss}_T$ search~\cite{Aaboud:2017iio} targets events with large missing transverse momentum and one or more displaced vertices 
with large track multiplicity (thus likely to correspond to hadronic decays). The analysis 
provides very detailed documentation\footnote{We note that CMS analyses 
searching for a events with an electron and a muon with large impact parameters~\cite{CMS:2016isf} are very well-documented 
and sensitive to the present scenario via decays $\chi_2 \to h \chi_1$ ($h \to e \nu_{e}\, \mu \nu_{\mu}$). 
The very small Higgs leptonic branching fraction ($\sim 0.4 \,\%$) results in a much weaker sensitivity w.r.t. DV + $E^{\rm miss}_T$ searches 
(see e.g.~\cite{Liu:2015bma}).}
for validation and recasting in its auxiliary material~\cite{ATLAS_SUSY_AUX}, and we give details on the analysis selection in Appendix~\ref{AppendixB}.
In order to validate our approach, we have applied the analysis to the model used by ATLAS for interpretation of their results, corresponding 
to pair-production of long-lived gluinos, with the gluino eventually decaying to a pair of SM quarks and a neutralino. 
The results of our validation are shown in Appendix~\ref{AppendixB}, which allow us to perform a recast of the ATLAS search in terms of 
the Higgs-assisted freeze-in DM scenario explored in this work (an analogous recasting analysis for this scenario has already been 
performed in~\cite{Calibbi:2018fqf}, 
with similar results).

\begin{figure}[h]

\centering

$\vcenter{\hbox{\includegraphics[width=0.7\textwidth]{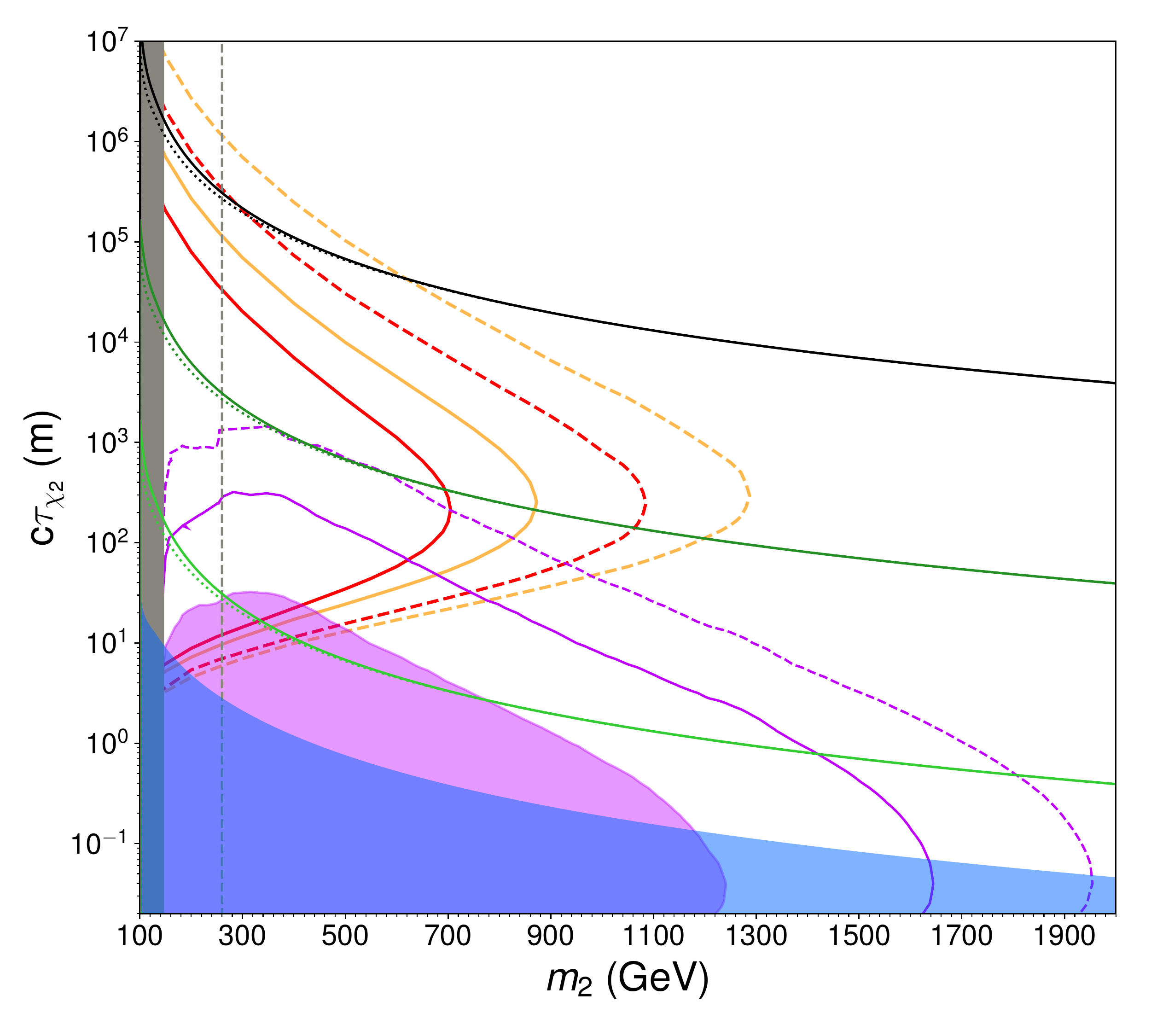}}}$
$\vcenter{\hbox{\includegraphics[width=0.29\textwidth]{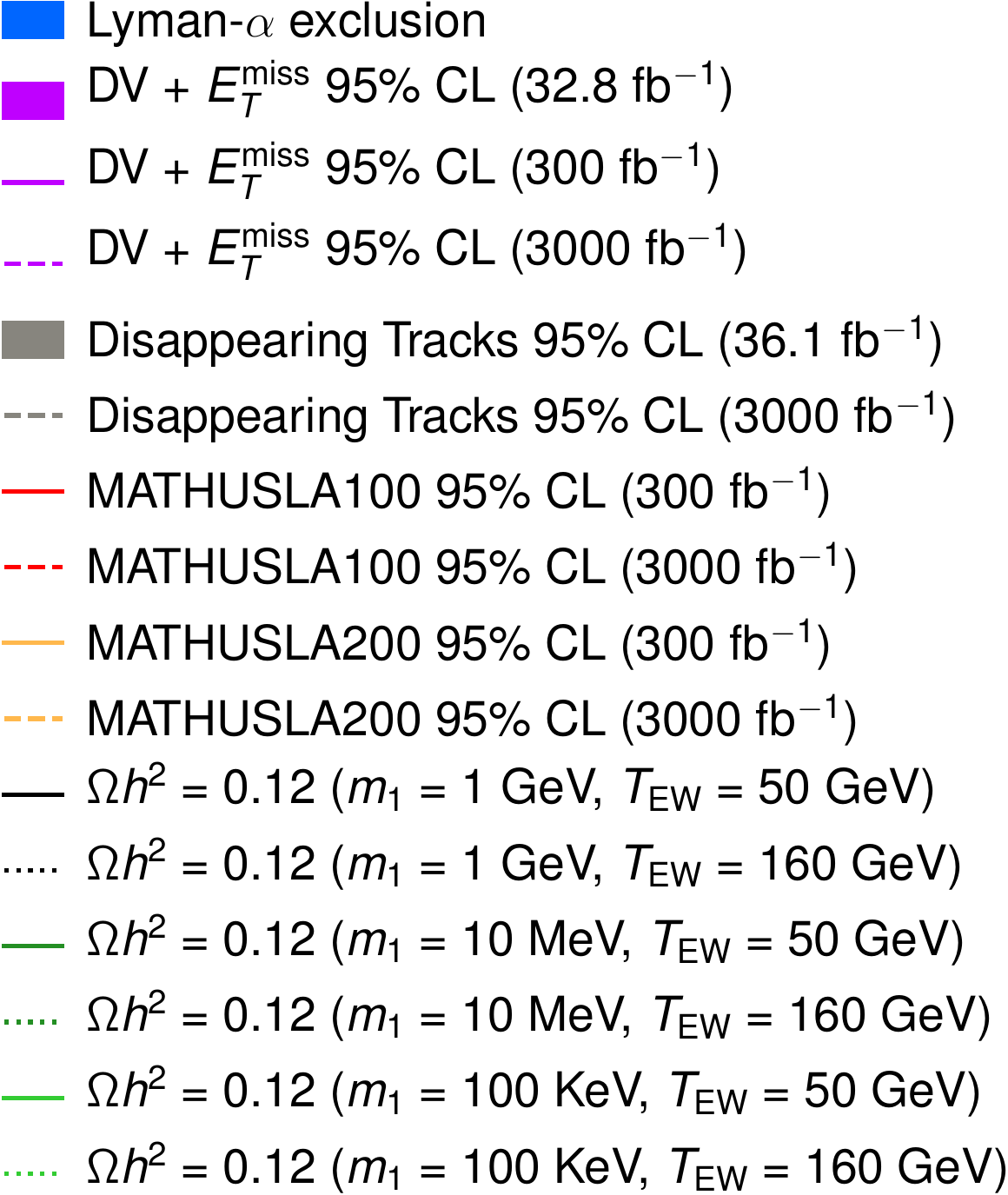}}}$

\caption{\small 13 TeV ATLAS DV $+$ $E^{\rm miss}_T$ search 95$\%$ C.L. exclusion limits with $32.8$ fb$^{-1}$ (purple region) in the ($m_2$, $c\tau_{\chi_2}$) plane, 
together with its extrapolation to 300 fb$^{-1}$ (solid purple line) and 3000 fb$^{-1}$ (dashed purple line). Also shown are the present 95\% C.L. exclusion limits from 
disappearing track searches (grey region) from section~\ref{sec:DT} and the projection for HL-LHC (vertical dashed grey line), as well as the MATHUSLA100 (red) and MATHUSLA200 (orange) projected 
95 $\%$ C.L. sensitivities for 300 fb$^{-1}$ (solid line) and 3000 fb$^{-1}$ (dashed line).  
Lines yielding 
the observed DM relic density $\Omega_{\mathrm{DM}} h^2 = 0.12$ (obtained with {\tt micrOMEGAs5.0}) 
are shown for $m_1 = 1$ GeV (black), $10$ MeV (dark green), $100$ KeV (light green), together with the bounds from 
Lyman-$\alpha$ (blue).
}
\label{Figure_FI_DV_MET}

\end{figure}

As in Section~\ref{MATHUSLA_Sec}, we use the Lagrangian implementation of the freeze-in DM model in {\tt FeynRules}~\cite{Alloul:2013bka}. 
We simulate Drell-Yan production $ p p \to \chi_2 \bar{\chi_2}$ with the decays of the 
$\chi_2\bar{\chi_2}$ pair to $Z \chi_1 \, Z \bar{\chi_1}$, $Z \chi_1 \,h \bar{\chi_1}$ or $h \chi_1\, h \bar{\chi_1}$
in {\tt Madgraph$\_$aMC@NLO}~\cite{Alwall:2014hca}. Since we find the kinematics 
of $\chi_2 \bar{\chi_2}$, $\chi_2 \psi^\pm$, and $\psi^+ \psi^-$ production to be very similar 
(both without and with the production of an extra hard jet, c.f. Figure~\ref{DM_mono_jet}), 
we normalize our signal cross section to the sum of respective LHC 
$\sqrt{s} = 13$ charged/neutral Higgsino production cross sections
at NLO+NLL obtained from {\tt Resummino-2.0.1}~\cite{Fuks:2012qx,Fuks:2013vua}.
After parton level event generation 
(we use the {\sc time\_of\_flight} option within {\tt Madgraph$\_$aMC@NLO} to introduce a non-zero displacement in the decay of $\chi_2$), 
our events are passed to  {\tt Pythia8}~\cite{Sjostrand:2014zea} which decays the $Z$ and $h$ bosons, and performs showering and hadronisation. Finally 
we cluster jets using the {\tt FastJet}~\cite{Cacciari:2011ma} implementation within {\tt Delphes~3}~\cite{deFavereau:2013fsa}.
Further details of our analysis are given in Appendix~\ref{AppendixB}.

The derived LHC 13 TeV 95\% C.L. limits from the ATLAS DV + $E^{\rm miss}_T$ search with $32.8$ fb$^{-1}$ 
on the ($m_2$, $c\tau_{\chi_2}$) parameter space of our present scenario
are shown in Figure~\ref{Figure_FI_DV_MET}, corresponding to a number of signal events $N_{\rm events} = 3$ (since the expected number of background events in the signal region 
is $0.02^{+0.02}_{-0.01}$, the current search may be considered as background-free).
We also show in Figure~\ref{Figure_FI_DV_MET} the 95\% C.L. sensitivity projection for the DV + $E^{\rm miss}_T$ search with 300 fb$^{-1}$ and 3000 fb$^{-1}$. 
In the former case, the search can still be regarded as background-free and we use $N_{\rm events} = 3$, whereas in the latter we expect $\sim 2$ background events, 
and the corresponding required number of signal events is found via the {\sl CLs} method \cite{Read:2002hq} to be $N_{\rm events} = 6$.
Finally, we also include in Figure~\ref{Figure_FI_DV_MET} the present and projected 95 $\%$ C.L. bounds from disappearing track searches (see section~\ref{sec:DT}) 
and the projected 95 $\%$ C.L. sensitivity
from MATHUSLA100 and MATHUSLA200 (see section~\ref{MATHUSLA_Sec}) to highlight the complementarity among the various searches. 
While for decay legths $c\tau_{\chi_2} \lesssim 100$ m 
the LHC DV + $E^{\rm miss}_T$ search yields the most sensitive probe of these scenario, for larger decay lengths MATHUSLA provides a major increase in sensitivity w.r.t. the main LHC detectors.

\section{Freeze-in at a $\sqrt{s} = 100$ TeV collider with a forward detector}
\label{100TeV_Sec}

As emphasized in the original MATHUSLA proposal~\cite{Chou:2016lxi}, a forward detector at a future proton-proton collider could 
significantly surpass the capabilities of the MATHUSLA surface detector proposal to search for very long-lived particles. 
In this section we analyze the sensitivity to the Higgs-mediated freeze-in DM scenario explored in this work
of a $\sqrt{s} = 100$ TeV hadron collider (from now on referred to as {\sl FCC-hh}) and a forward LLP detector with volume 
given by $z \in [20, \,40]$ m, $\rho = \sqrt{x^2 + y^2} \in [5, \,30]$ m as measured from the {\sl FCC-hh} interaction point. 

\begin{figure}[t]

\centering

$\vcenter{\hbox{\includegraphics[width=0.7\textwidth]{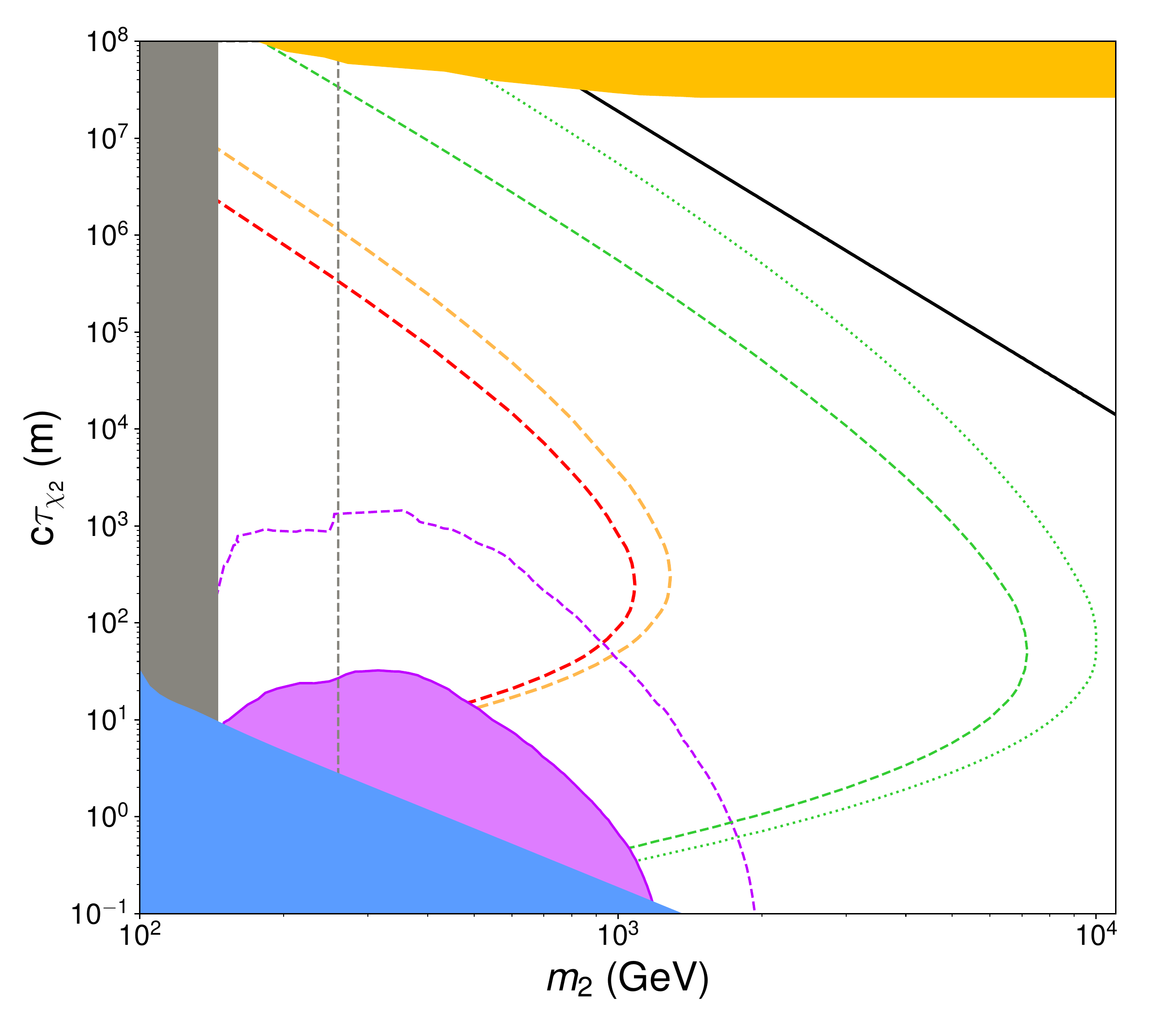}}}$
$\vcenter{\hbox{\includegraphics[width=0.29\textwidth]{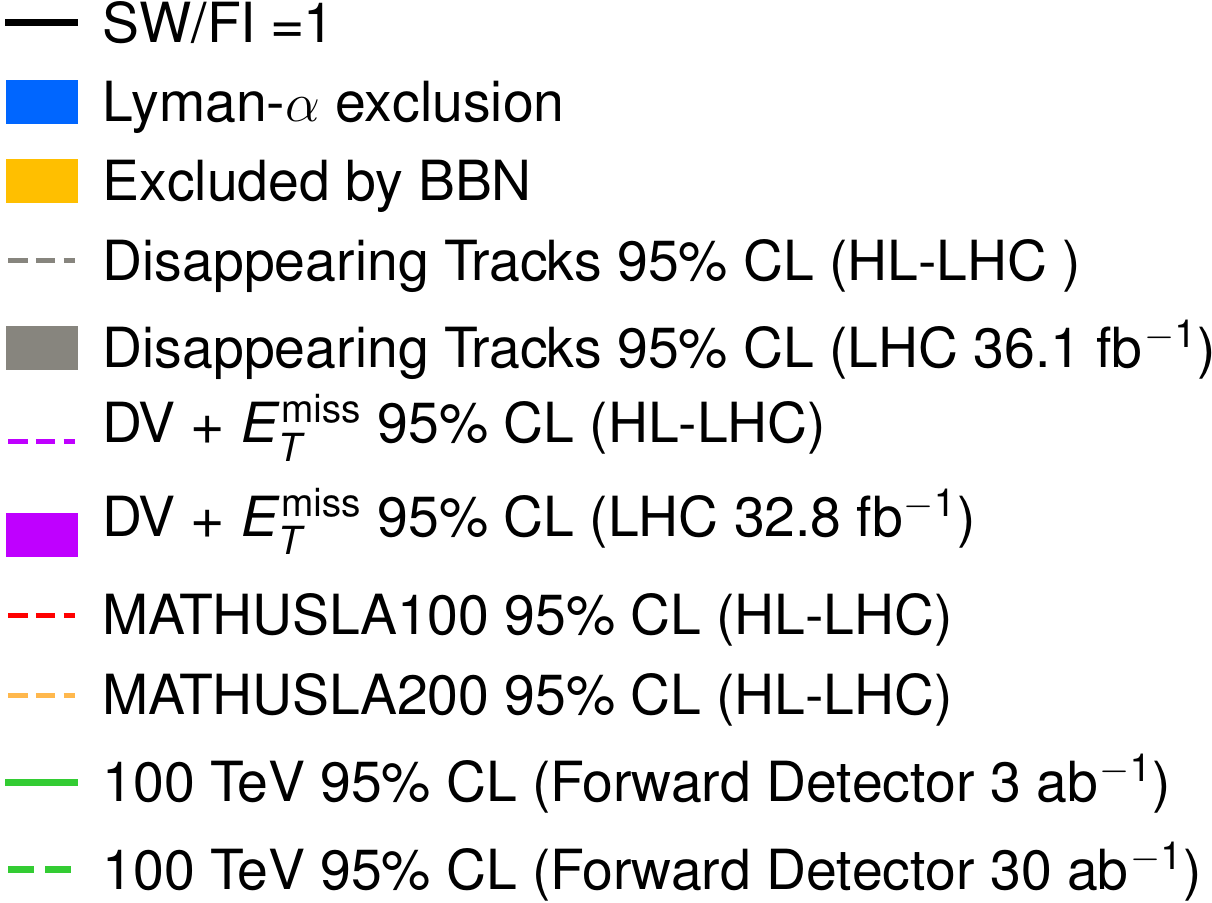}}}$

\vspace{-2mm}

\caption{\small 
{\sl FCC-hh}~forward detector 95\% C.L. sensitivity projections ($N_{\mathrm{events}} = 3$) for $\mathcal{L} = 3$ ab$^{-1}$ (dashed-green line) and $\mathcal{L} = 30$ ab$^{-1}$ (dotted-green line). Shown for comparison are the MATHUSLA100 (dashed-red line) and MATHUSLA200 (dashed-orange line) sensitivity projections with $\mathcal{L} = 3$ ab$^{-1}$  from section~\ref{MATHUSLA_Sec}, the present
13 TeV ATLAS DV $+$ $E^{\rm miss}_T$ search exclusion limits (purple region) and the extrapolation to HL-LHC (dashed-purple line) from section~\ref{sec:DV}, as well as the present exclusion limits from disappearing track searches (grey region) and their extrapolation to HL-LHC (vertical dashed-grey line) from section~\ref{sec:DT}. Also shown are the bounds from Lyman-$\alpha$ (blue region) and BBN (yellow region), and the boundary of the super-WIMP parameter-space region (solid-black line), given by SW/FI$ = 1$ (recall Figure~\ref{superWIMP} and the discussion in section~\ref{sec:DM_Relic_SW}).
}
\label{Figure_FI_100TeV}

\end{figure}

We simulate Drell-Yan production $ p p \to \chi_2 \bar{\chi_2}$, $ p p \to \bar{\chi_2} \psi^{-}$, 
$p p \to \chi_2 \psi^{+}$, $p p \to \psi^{+}\psi^{-}$
at $\sqrt{s} = 100$ TeV in {\tt Madgraph$\_$aMC@NLO}, normalizing the respective cross sections to the corresponding 
NLO+NLL pure Higgsino production cross sections computed with {\tt Resummino-2.0.1}, and shown in Appendix~\ref{AppendixC}. The decays of the charged states into $\chi_2$ and $\bar{\chi_2}$ 
are again considered to yield neutral states with approximately identical kinematics to that of the decaying charged states.
We follow the same procedure used in section~\ref{MATHUSLA_Sec} for MATHUSLA to estimate the event yield at the forward detector, computing from simulation the number of signal events $N_{\mathrm{events}}$ 
from the probability for a particle $\chi_2$ to decay within the forward detector volume (we assume perfect detector performance as well as a 
background-free environment), given by~\eqref{e.Pdecay}.  
The forward detector geometric acceptance for our signal  
is found from simulation to be $\epsilon_{\mathrm{geometric}} \sim 0.5$, which is roughly ten times larger than the one of MATHUSLA due 
to the rather forward nature of our Drell-Yan events for $\sqrt{s} = 100$ TeV proton-proton collisions.

In Figure~\ref{Figure_FI_100TeV} we show our {\sl FCC-hh} forward detector 95\% C.L. sensitivity projections, given by $N_{\mathrm{events}} = 3$, for two choices of {\sl FCC-hh} integrated luminosity, $\mathcal{L} = 3$ ab$^{-1}$ and $\mathcal{L} = 30$ ab$^{-1}$. 
The mass reach shows a dramatic increase w.r.t. MATHUSLA and HL-LHC\footnote{Obviously, a DV $+$ $E^{\rm miss}_T$ search at {\sl FCC-hh} would significantly increase the mass reach of the corresponding search at HL-LHC. Still, already for $c\tau_{\chi_2} \gtrsim 1$ m the forward detector is expected to be more sensitive than the main {\sl FCC-hh} detector(s), as the former will have much less background and the geometric acceptances of both are comparable (as opposed to HL-LHC vs MATHUSLA, case in which the smaller geometric acceptance $\epsilon_{\mathrm{geometric}}$ of MATHUSLA results in MATHUSLA being more sensitive than the LHC DV $+$ $E^{\rm miss}_T$ search only for $c\tau_{\chi_2} \gtrsim 100$ m).}, being able to probe LLPs with $c\tau_{\chi_2} \sim 50$ m up to $m_2 \sim 10$ TeV. Figure~\ref{Figure_FI_100TeV} also shows that the forward detector would be able to probe lifetimes up to the BBN bound for $m_2 \lesssim 600$ GeV, and approach the boundary of the super-WIMP parameter space region (solid black line in Figure~\ref{Figure_FI_100TeV}).

\section{Conclusions}
\label{sec_conclusions}

Freeze-in constitutes a well-motivated and appealing mechanism for DM production in the early Universe, yet challenging to probe experimentally given the feeble interactions between DM and the visible sector. In this sense, collider searches for LLPs have been recently regarded as possible probes of DM freeze-in production 
via the decay of a parent particle which is accessible at colliders.
In this work we have studied the prospects of probing DM freeze-in from the decay of neutral parent particles which belong to the thermal bath, through LLP signatures. 
Using as a case-study a model featuring Higgs-associated LLP decays, 
we have obtained the experimental sensitivity of the proposed MATHUSLA surface detector and ATLAS/CMS at HL-LHC to the cosmologically relevant 
freeze-in DM parameter space. Our results show the high degree of complementarity between MATHUSLA and the main LHC detectors, together being able to 
probe DM masses from the Lyman-$\alpha$ bound (of a few keV) up to masses of few GeV, and neutral parent particle masses up to $\sim 2$ TeV. 
Besides, for parent particle masses of $\mathcal{O}(100)$~GeV, MATHUSLA could probe lifetimes close to the BBN limit.

We have also analyzed the improvement in sensitivity that would come from a forward LLP detector within a future 100 TeV hadron collider ({\sl FCC-hh}),
finding that such an improvement would be quite dramatic and would allow to probe parent LLPs with masses up to $10$ TeV for DM masses in the MeV range, 
as well as reach the BBN limit for parent LLP masses below 600 GeV.
    
\vspace{2mm}
    
In addition to the above results, we have discussed in some detail several technical aspects of freeze-in DM scenarios: the first aspect, being particular to our Higgs-assisted freeze-in DM model, concerns the EW phase transition, which switches-on some DM production processes when it takes place and plays a key role in regularising certain $2 \to 2$ scattering processes which would otherwise make DM production 
UV-sensitive. 
At the same time, our calculation of the DM production prior to the EW phase transition includes the corrections due to the thermal mass of the Higgs doublet and the parent particle. 
The inclusion of thermal masses in DM relic density computations is a rather generic issue that has so far been disregarded for freeze-in via decay scenarios\footnote{For freeze-in via scattering, see the very recent studies
~\cite{Heeba:2019jho,Hambye:2019dwd,Darme:2019wpd}. Note that the model in \cite{Darme:2019wpd} produces a pair of DM particles from the decay of the parent particle, contrary to our scenario which is the ``standard'' freeze-in via decay, where only a single DM particle is produced along with a thermal bath particle.} and can have a significant quantitative impact on the results. 
The other technical aspects analyzed also apply more generally to models of freeze-in DM production from decay, and correspond to the impact of $2\to 2$ scattering processes on the DM relic abundance and the interplay between freeze-in and super-WIMP Dark Matter production mechanisms.

\acknowledgments

We would like to thank Thomas Konstandin, Alberto Mariotti, Laura Lopez-Honorez, Felix Kahlhoefer, Saniya Heeba, Steven Lowette, Suzanne Westhoff and Manimala Chakrabourti for useful discussions. We also thank Benjamin Fuks for pointing us to the use of {\tt Resummino}. J.M.N. is supported by the Programa Atracci\'on de Talento de la Comunidad de Madrid via grant 2017-T1/TIC-5202. 
B.Z. acknowledges support from the Programa Atraccion de Talento de la Comunidad de Madrid under grant n. 2017-T2/TIC-5455.
P.T. acknowledges support from the Deutsche Forschungsgemeinschaft (DFG) through the Emmy Noether Grant No. KA 4662/1-1 and the Collaborative Research Center TRR 257 
``Particle Physics Phenomenology after the Higgs Discovery''.
J.M.N. and B.Z. acknowledge support from the Spanish MINECO's ``Centro de Excelencia Severo Ochoa" Programme via grant SEV-2016-0597.   
J.M.N. and P.T. were supported by the European Research Council under the European Union’s Horizon 2020 program, ERC Grant Agreement 648680 (DARKHORIZONS) 
during the early stages of this work. J.M.N. thanks the Korean Institute for Advanced Study (KIAS) for hospitality during completion of this work.

\appendix 
\section{$1\to 2$ vs  $2 \to 2$ freeze-in processes}\label{AppendixA}
\setcounter{equation}{0}
\renewcommand{\theequation}{A.\arabic{equation}}

\begin{figure}[h]
\begin{center}
\includegraphics[width=0.36\textwidth]{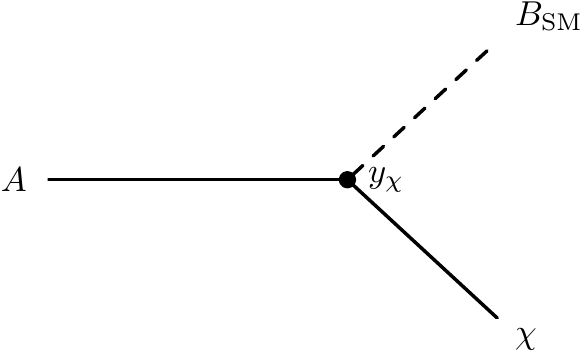}
\hspace{15mm}
\includegraphics[width=0.36\textwidth]{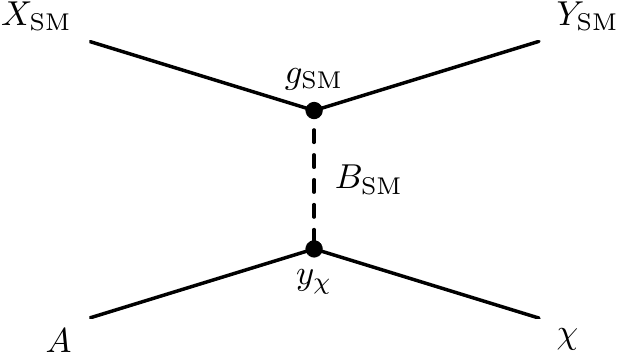}
\vspace{-3mm}
\caption{\small Feynman diagrams for the processes $A \to B_{\rm SM} \chi$ (left) and $X_{\rm SM} + A \to Y_{\rm SM} + \chi$ (right).
}
\label{Fig_Feynman_12_22}
\end{center}
\end{figure}

As outlined in Section~\ref{subsec:DM_Relic_EWSB}, whenever there exists a decay $A \to B_{\rm SM} \chi$ contributing to freeze-in DM production, there will be 
related $2 \to 2$ scattering processes also contributing to DM freeze-in at the same order in the freeze-in coupling $y_{\chi}$, where the SM 
particle $B_{\rm SM}$ appears in the $t$-channel of the scattering process 
$X_{\rm SM} + A \to Y_{\rm SM} + \chi$, with both $X_{\rm SM}$ and $Y_{\rm SM}$ SM particles. 
This is shown in Figure~\ref{Fig_Feynman_12_22}, with the SM coupling among $X_{\rm SM}$, $Y_{\rm SM}$ and $B_{\rm SM}$ denoted generically by $g_{\rm SM}$.

We define a measure of the relative importance of these $2 \to 2$ scattering processes w.r.t. the $1 \to 2$ decays $A \to B_{\rm SM} \chi$, simply given by 
the ratio of DM yields from $1 \to 2$ and $2 \to 2$ processes (with thermal corrections neglected in this discussion):
\begin{eqnarray}
\label{yield_ratio}
\frac{Y_{2\to2}}{Y_{1\to2}} &=& \frac{1}{48\, \pi^3 m_A^2 \Gamma_A} \\  &\times& 
\int_0^{\infty} x^3 \, dx \int_{(m_A + m_{X})^2}^{\infty} \sqrt{s} \,\left[s - (m_A + m_{X})^2\right] 
K_1\left(\frac{\sqrt{s}\, x}{m_A + m_{X}} \right)\, \sigma(s) \, ds \nonumber \\
 &=& \frac{1}{32\, \pi^2 m_A^2 \Gamma_A} \times \int_{(m_A + m_{X})^2}^{\infty} \frac{(m_A+m_X)^4\, \sqrt{s} \,\left[s - (m_A + m_{X})^2\right]}{s^2} \, \sigma(s) \, ds \nonumber
\end{eqnarray}
with $\Gamma_A$ the decay width of the process $A \to B_{\rm SM} \chi$, $\sigma(s)$ the cross section for the 
process $X_{\rm SM} + A \to Y_{\rm SM} + \chi$, $m_X$ and $m_A$ the mass of the initial states $A$ and $X_{\rm SM}$ and $K_1$ the first 
modified Bessel function of the 2$^{\rm nd}$ kind. In the last step of~\eqref{yield_ratio} we have performed explicitly the integration in $x = m_A/T$
(only possible when thermal corrections are neglected). 

\begin{figure}[t]
\begin{center}
\includegraphics[width=0.85\textwidth]{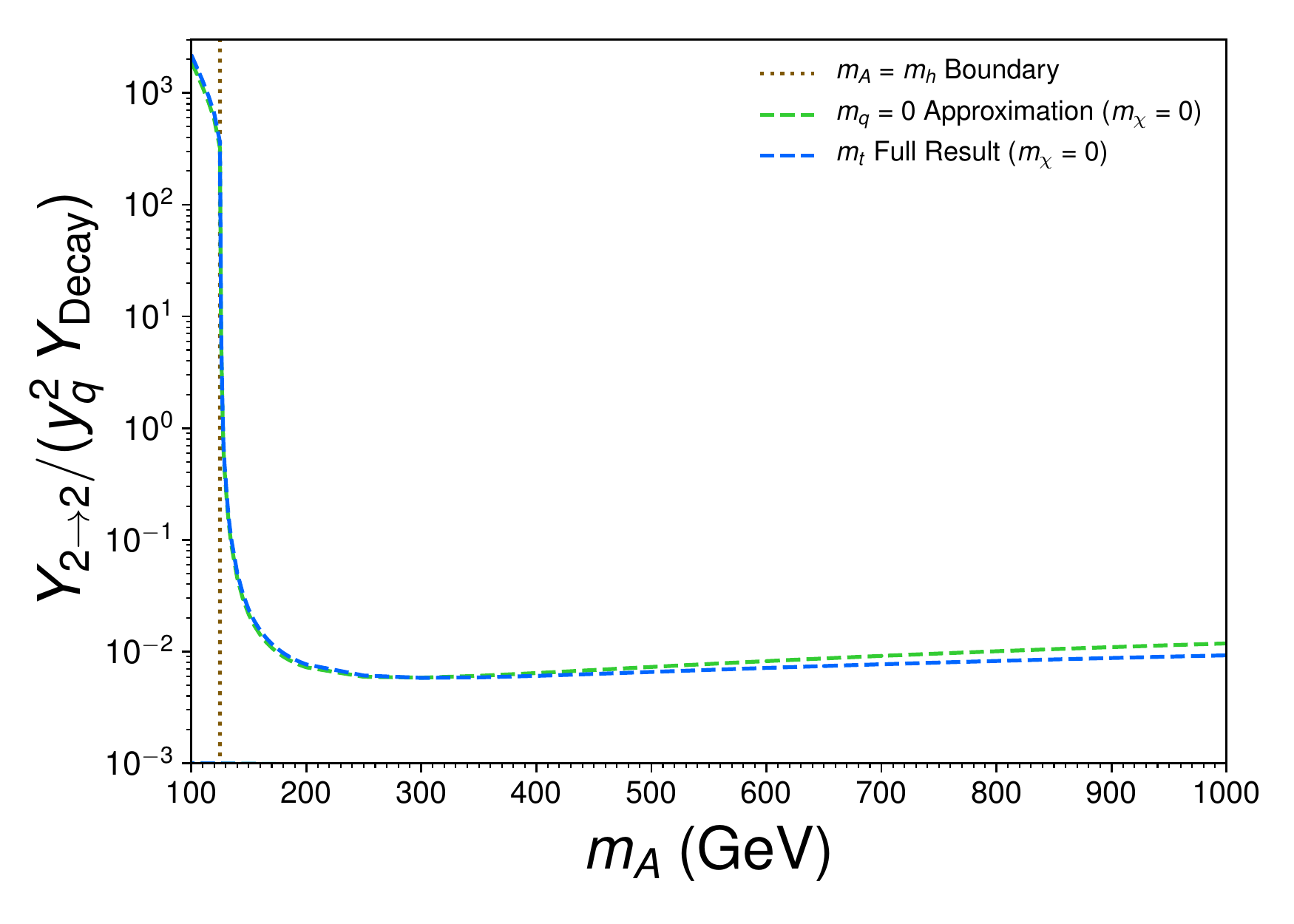}
\vspace{-3mm}
\caption{\small Ratio of decays ($1 \to 2$ for $m_A > m_h$ and $1 \to 3$ for $m_A < m_h$) to $2 \to 2$ freeze-in DM production processes $Y_{2\to 2}/(y^2_q\,Y_{\mathrm{Decay}})$ as a function of $m_A$ and for $m_{\chi} =0$, neglecting $m_q$ (green) and including the full $m_t$ dependence (blue). The boundary $m_A = m_h$ (below which the $1\to 2$ decay cannot happen) is shown as a brown dotted vertical line.}
\label{Fig:yield_ratio}
\end{center}
\end{figure}

As discussed above, $\Gamma_A$, $\sigma(s) \propto y_{\chi}^2$ and so the ratio~\eqref{yield_ratio} is independent of $y_{\chi}$ and might turn out to be sizable. 
In the following, we compute the above ratio in our freeze-in scenario for $B_{\rm SM} = h$ and $X_{\rm SM} = Y_{\rm SM} = q$, a SM quark with Yukawa coupling $y_q$ 
(other possibilities for $X_{\rm SM}$ and $Y_{\rm SM}$ such as SM gauge bosons $W$, $Z$ would yield similar results in this case, being the nature of the $t$-channel mediator the relevant ingredient here). 
The cross section $\sigma(s)$ has a simple analytic form in the limit $m_q,\,m_{\chi} \to 0$
\begin{equation}
\label{XS_vs_width}
\frac{\sigma(s)}{y_{\chi}^2} =  \frac{y_q^2}{64\,\pi} \left[ 
\frac{(s - 2\,m_A^2 + 2\,m_h^2)}{(s -m_A^2)\,(s -m_A^2 + m_h^2)} + \frac{(m_A^2 - 2\, m_h^2)}{(s -m_A^2)^2} \, 
\mathrm{log}\left(\frac{s- m_A^2 + m_h^2}{m_h^2} \right)\right] \,.
\end{equation}
Inserting~\eqref{XS_vs_width} into~\eqref{yield_ratio} and with $\Gamma_A$ given by $\Gamma(\chi_2 \to h\chi_1)$ in  Eq.~\eqref{decayFIDM_1}, we obtain 
the ratio $Y_{2\to 2}/(y^2_q\,Y_{1\to 2})$ as a function of $m_A$, shown in Figure~\ref{Fig:yield_ratio}.
We nevertheless note that for $q$ being the top quark, which corresponds to the leading $2\to 2$ scattering process (since $\sigma(s) \propto y^2_{t}$ 
in that case), the top mass $m_t$ cannot be neglected. In this case we obtain the cross section $\sigma(s)$ numerically 
and show the ratio $Y_{2\to 2}/(y^2_t\,Y_{1\to 2})$ as a function of $m_A$ in Figure~\ref{Fig:yield_ratio}, keeping for simplicity $m_{\chi} \to 0$. It becomes clear that the decay 
processes are largely dominant except in the limit $m_A \to m_h$, 
and for $m_A < m_h $ when $2$-body decays are forbidden and the ratio~\eqref{yield_ratio} should instead be understood as $Y_{2\to2}/Y_{1\to 3} $ (the relevant decays of $A$ are 3-body). In Figure~\ref{Fig:yield_ratio} we also show the region $m_A < m_h$ down to 
$m_A = 100$ GeV, where the ratio $Y_{2\to 2}/(y^2_q\,Y_{1\to 3})$ is very large and $2 \to 2$ scatterings clearly dominate freeze-in DM production.  

\vspace{2mm}

The case where $B_{\rm SM}$ is an EW gauge boson nevertheless behaves in a different fashion to the above, and is worth discussing separately.
Focusing on $B_{\rm SM} = Z$ (the analysis for $B_{\rm SM} = W^{\pm}$ is analogous), we note that the 
cross section for the $2 \to 2$ scattering process $X_{\rm SM} + A \to Y_{\rm SM} + \chi$ (with $X_{\rm SM} = Y_{\rm SM} = q$, a SM quark) 
mediated by a $t-$channel $Z$ boson is given in the limit $m_q,\,m_{\chi} \to 0$ by
\begin{eqnarray}
\label{XS_vs_width_Z}
\frac{\sigma(s)}{y_{\chi}^2} =  \frac{m_Z^4\, (5 - 4 \, \mathrm{cos}^2 \,\theta_W + 
8 \, \mathrm{cos}^4 \,\theta_W)}{144\,\pi\,m_A^2\,v^2} \times \quad \quad \quad \quad 
\quad \quad \quad \quad \quad \quad \quad \quad \quad \quad \quad \quad \quad \quad \\ \left[ 
\frac{\left[s \left( s + m_Z^2/2\right) + (s + m_Z^2) (m_Z^2 - m_A^2)\right]}{m_Z^2\,(s -m_A^2)\,(s -m_A^2 + m_Z^2)} - \frac{s + m_Z^2 - m_A^2/2}{(s -m_A^2)^2} \, 
\mathrm{log}\left(\frac{s- m_A^2 + m_Z^2}{m_Z^2} \right)\right] \,. \nonumber
\end{eqnarray}
and scales as $m_A^{-2}$ (independent of $s$) for $s \gg m_A^2, m_Z^2$, as opposed to~\eqref{XS_vs_width} which scales as $s^{-2}$ in that limit.
Naively, this would lead to a UV divergent result for~\eqref{yield_ratio} (without considering thermal corrections, which are crucial in this case), signaling a UV dominated freeze-in production mechanism from  
$2 \to 2$ scattering processes mediated by a $t-$channel $Z$ boson. 
We note however that the interaction among 
$A$, $\chi$ and the EW gauge bosons vanishes when the EW symmetry is restored above the EW phase transition temperature (recall the discussion in section~\ref{subsec:DM_Relic_EWSB}), which regulates the 
integral $\int \, ds$ in~\eqref{yield_ratio} and
again yields a subdominant contribution from $2 \to 2$ scattering processes to the freeze-in DM abundance.

Finally, it is worth stressing that in a general scenario (such as ours) with various $1\to 2$ and $2\to 2$ processes contributing to 
DM freeze-in, the accurate expression analogous to~\eqref{yield_ratio} may not be as simple, with different contributions 
becoming relevant at different temperatures (particularly due to electroweak symmetry breaking in our scenario).

\section{$13$ TeV ATLAS DV + $E^{\rm miss}_T$ search recast} 
\label{AppendixB} 
\setcounter{equation}{0}
\renewcommand{\theequation}{B.\arabic{equation}}
 
We describe here our recast of the ATLAS search for displaced vertices plus missing transverse energy~\cite{Aaboud:2017iio}
at $\sqrt{s} = 13$ TeV with $32.8$ fb$^{-1}$. The auxiliary material from the ATLAS search~\cite{ATLAS_SUSY_AUX} provides 
efficiencies that can be applied to simulated truth level event samples.
The search defines tracks (which will be associated to displaced vertices DV below) as:
\begin{itemize}
    \item The particle associated to the track is charged and stable.
    \item The particle has $p_T > 1$~GeV.
    \item The particle has a transverse impact parameter $d_0 > 2$~mm.
\end{itemize}
Displaced vertices are constructed from these tracks. The DV must satisfy:
\begin{itemize}
    \item 4~mm~$ < R < $~300~mm, where $R = \sqrt{x^2 + y^2}$ is the transverse distance to the interaction point, and $|z| <$ 300 mm.
    \item The DV must have 5 or more tracks (as defined above) associated to it.
    \item The DV mass $m_{\mathrm{DV}}$ must be at least 10~GeV, calculated from the three-momenta of the tracks (as defined above) associated to it and assuming 
    a charged pion mass for each of the tracks.
\end{itemize}

\begin{figure}[t]
    \centering
    \includegraphics[width=0.84 \textwidth]{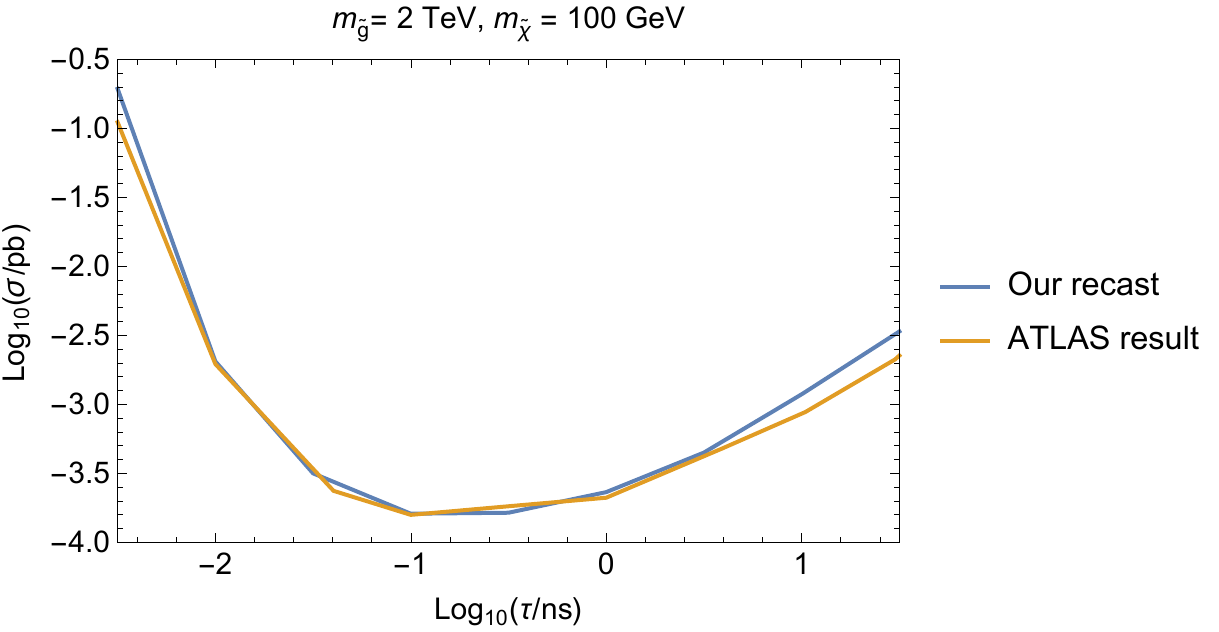}
    \vspace{2mm}
    \includegraphics[width=0.84 \textwidth]{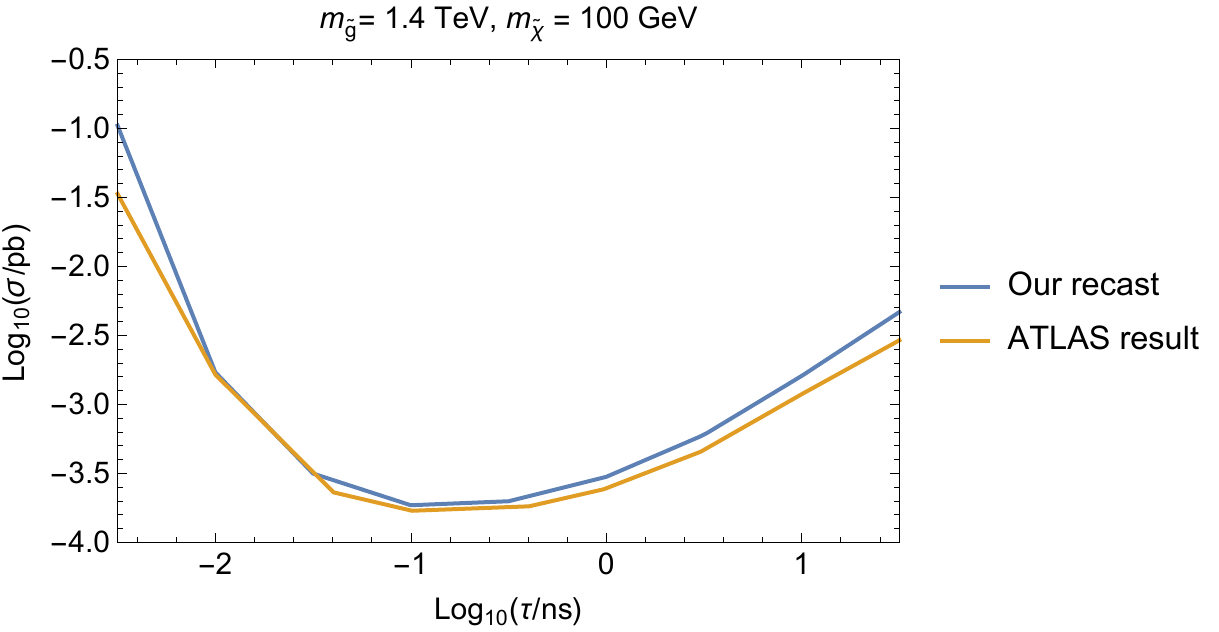}
    \caption{Excluded cross sections at 95\% C.L. as a function of the gluino lifetime, for a long lived gluino decaying to a quark pair 
    and a neutralino, compared to the published result from ATLAS~\cite{Aaboud:2017iio}. The gluino mass is fixed to 2~TeV (top) or 1.4~TeV (bottom). 
    The neutralino mass is fixed to 100~GeV.}
    \label{fig:ValidationLifetime}
\end{figure}

\noindent Finally, the whole event is required to have:
\begin{itemize}
    \item $E^{\rm miss}_T > 200$~GeV.
    \item 75\% of the events should have at least one jet\footnote{The ATLAS analysis requires trackless jets, where a trackless jet is one where the scalar 
    sum of the charged particle $p_T$ is less than 5~GeV for those particles with small impact parameter with respect to the primary vertex. However, 
    since the small impact parameter is not defined by ATLAS, we are unable to impose the trackless requirement in our recast.} with $p_T > 70$~GeV or at least 
    two jets with $p_T > 25$~GeV.
    \item At least one displaced vertex (as defined above).
\end{itemize}

\begin{figure}[t]
    \centering
    \includegraphics[width=0.84 \textwidth]{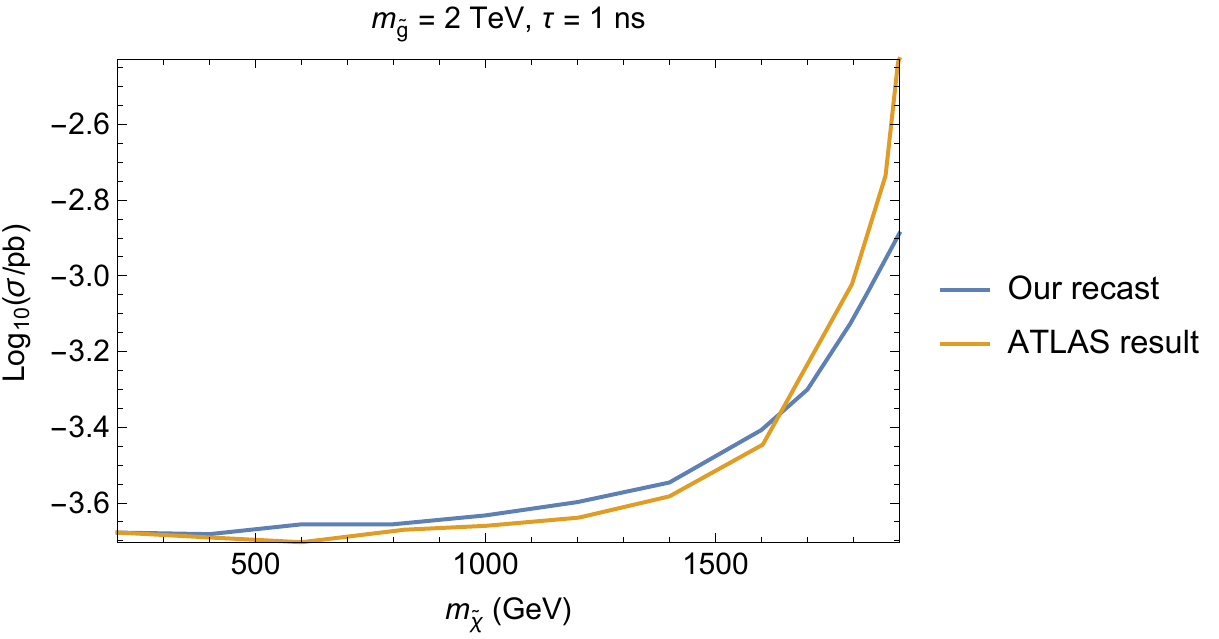}
        \vspace{2mm}
    \includegraphics[width=0.84 \textwidth]{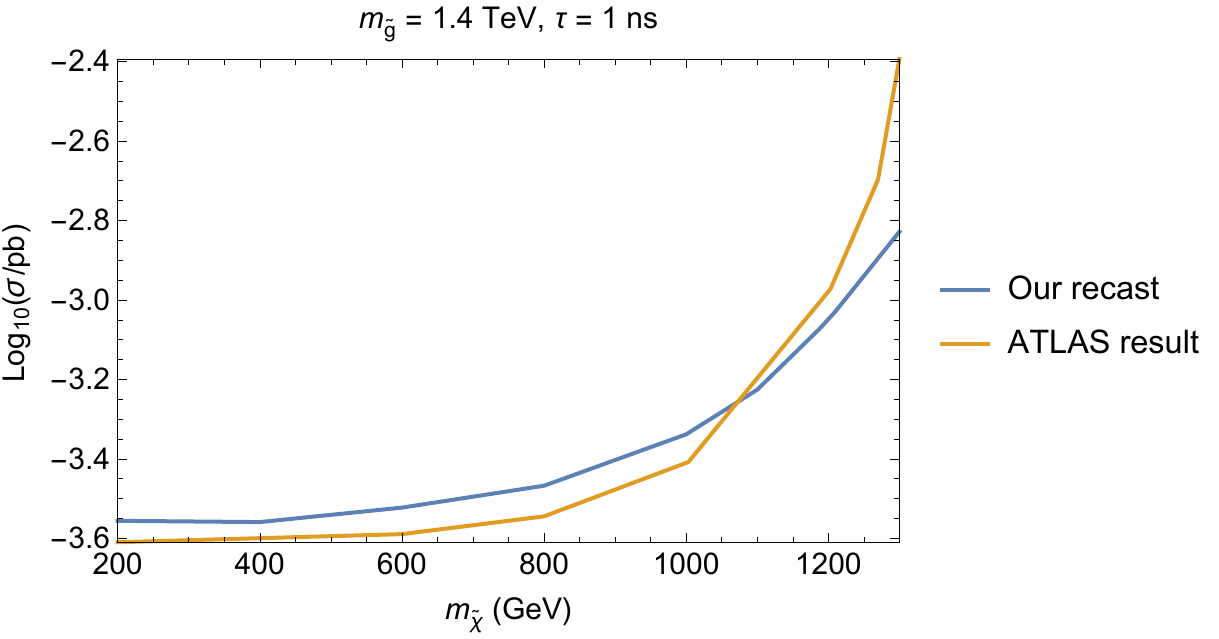}
    \caption{Excluded cross sections at 95\% C.L. as a function of the neutralino mass, for a long lived gluino decaying to a quark pair 
    and a neutralino, compared to the published result from ATLAS~\cite{Aaboud:2017iio}. The gluino mass is fixed to 2~TeV (top) or 1.4~TeV (bottom). 
    The gluino lifetime is fixed to 1~ns.}
    \label{fig:ValidationMass}
\end{figure}

After our signal event simulation with {\tt Madgraph$\_$aMC@NLO}~\cite{Alwall:2014hca} and {\tt Pythia8}~\cite{Sjostrand:2014zea}, 
followed by jet-clustering using the {\tt FastJet}~\cite{Cacciari:2011ma} implementation within {\tt Delphes~3}~\cite{deFavereau:2013fsa}, we 
perform the track, vertex, and event selection using the {\sc ROOT}~\cite{Antcheva:2009zz} output of {\tt Delphes~3}. For those 
DV in events that pass all the cuts, we apply the vertex-level and event-level efficiencies provided within the auxiliary 
material of the ATLAS search, obtaining overall event selection efficiencies for 
$Z Z + E^{\rm miss}_T$, $Z h + E^{\rm miss}_T$ and $h h + E^{\rm miss}_T$ signal decay channels, respectively.
These overall efficiencies for the three separate channels are a function of the $\chi_2$ mass and lifetime, 
and we therefore generate events 
for the three different different decay channels and different $\chi_2$ masses and decay lengths\footnote{In practice, 
we have used a fixed $c \tau_{\chi_2} = $~0.1 m for event generation, noting that since each 
displacement variable is proportional to $c \tau_{\chi_2}$, we can trivially rescale the position space coordinates of each particle to different values of 
$c \tau_{\chi_2}$, before applying the cuts listed above.} $c\tau_{\chi_2}$.

In order to validate the above analysis, we have applied our recast to a long-lived gluino model, where the gluino decays to a quark pair and 
a neutralino. This is the model used by ATLAS to interpret the results from their DV + $E^{\rm miss}_T$ search. 
We show our derived 95\% C.L. exclusion limits, as compared to those shown by ATLAS~\cite{Aaboud:2017iio}, in Figure~\ref{fig:ValidationLifetime} as a function 
of the gluino lifetime and in Figure~\ref{fig:ValidationMass} as a function of the neutralino mass. We find overall good agreement, except in the 
compressed region (neutralino mass approaching the gluino mass) where our derived limits are stronger than those of ATLAS, 
signaling that our recasting procedure is less trustworthy for a compressed scenario. In our freeze-in DM model, 
this happens when the mass of $\chi_2$ approaches the SM Higgs mass (the DM candidate $\chi_1$ is in general approximately massless on LHC scales), 
such that our derived limits in that region will be stronger than the actual, would-be ATLAS limits.


\begin{figure}[h]
    \centering
    \includegraphics[width=0.73 \textwidth]{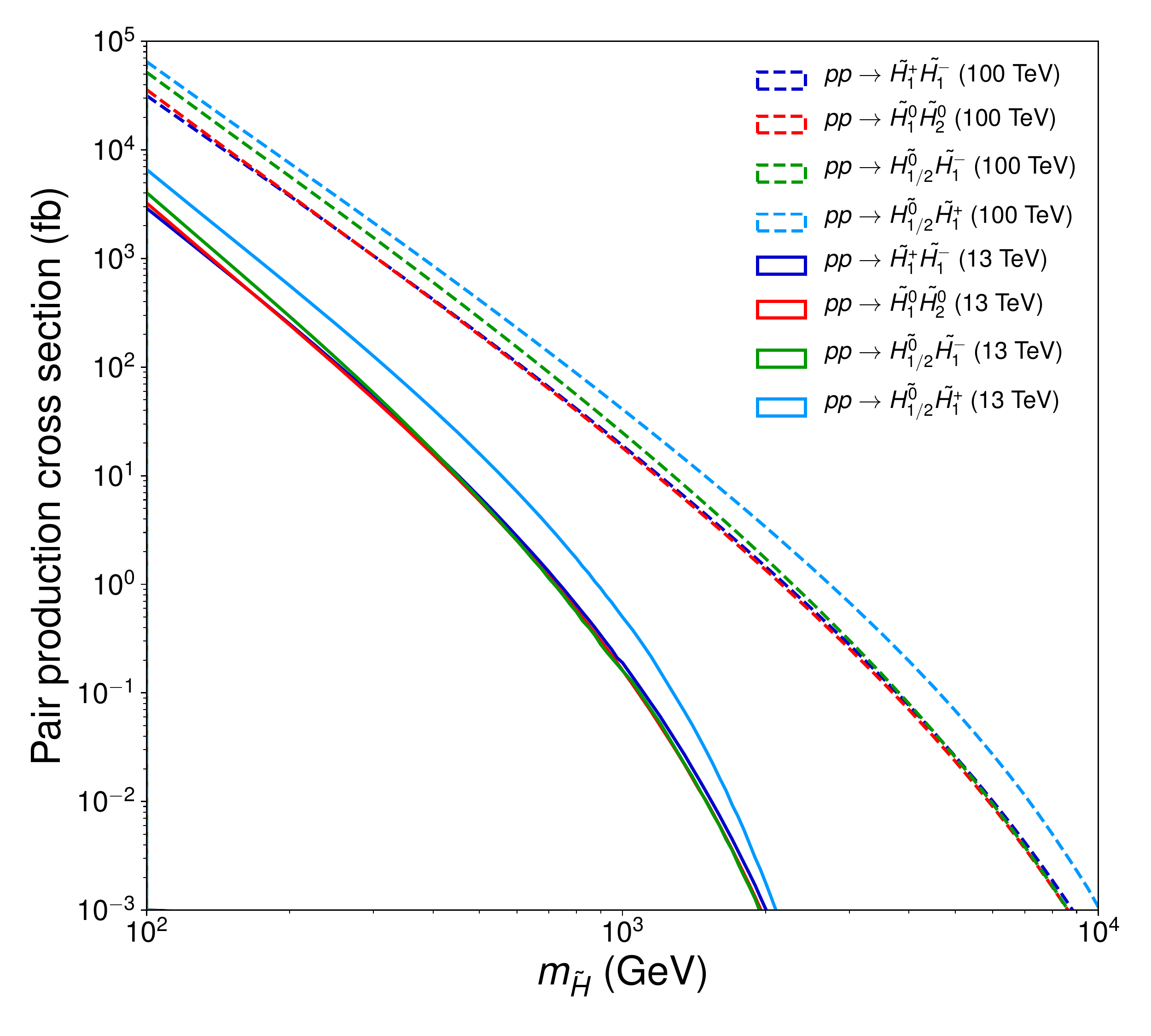}
    
    \vspace{-3mm}
    
    \caption{13 TeV (solid) and 100 TeV (dashed) NLO+NLL pure Higgsino pair production cross sections (assuming decoupled squarks) as a function of the Higgsino mass $m_{\tilde{H}}$ obtained with {\tt Resummino-2.0.1} with the PDF set {\sc MSTW2008nlo90cl}. The processes shown are $ p p \to \tilde{H^{0}_1}\tilde{H^{0}_2}$ (red), $ p p \to \tilde{H^{0}_{1/2}} \tilde{H^{-}_1}$ (green), 
$p p \to \tilde{H^{0}_{1/2}} \tilde{H^{+}_1}$ (light blue) and $p p \to \tilde{H^{+}_1} \tilde{H^{-}_1}$ (dark blue).}
    \label{fig:Higgsino_XS}
\end{figure}

\section{Higgsino pair production cross sections at $\sqrt{s} = 13,\,100$ TeV} 
\label{AppendixC} 
\setcounter{equation}{0}
\renewcommand{\theequation}{C.\arabic{equation}}

In this Appendix, we show the NLO+NLL pure Higgsino pair production cross sections (assuming decoupled squarks) used in this work, corresponding to 
$ p p \to \tilde{H^{0}_1}\tilde{H^{0}_2}$, $ p p \to \tilde{H^{0}_{1/2}} \tilde{H^{-}_1}$, 
$p p \to \tilde{H^{0}_{1/2}} \tilde{H^{+}_1} $ and $p p \to \tilde{H^{+}_1} \tilde{H^{-}_1}$. These are computed as a function of the 
Higgsino mass $m_{\tilde{H}} \equiv m_{\tilde{H^{\pm}_1}} = m_{\tilde{H^{0}_1}} = m_{\tilde{H^{0}_2}}$ 
from {\tt Resummino-2.0.1}~\cite{Fuks:2012qx,Fuks:2013vua} with the PDF set {\sc MSTW2008nlo90cl} from {\tt LHAPDF}~\cite{Buckley:2014ana}, both 
at $\sqrt{s} = 13$ TeV and $\sqrt{s} = 100$ TeV, and shown in Figure~\ref{fig:Higgsino_XS}. The $\sqrt{s} = 100$ TeV cross sections are found to be in qualitative agreement with the LO pure Higgsino pair production cross sections computed in~\cite{Acharya:2014pua}.

\bibliography{MATHUSLA}{}
 
\bibliographystyle{JHEP}

 \end{document}